\documentclass[11pt,noindent]{article}

\usepackage{graphics,cite,epsfig,float,tablefootnote}
\usepackage{amsfonts,xfrac,slashed}
\usepackage{amstext,amssymb,amsfonts}
\usepackage{amsmath,mathbbol}
\usepackage{a4,graphicx}
\usepackage{dcolumn}
\usepackage{color}
\usepackage{enumerate}

\usepackage{array}
\usepackage{multirow}
\usepackage{ulem}
\usepackage{soul}
\usepackage{bm}
\usepackage[utf8]{inputenc}

\begin{document}
\begin{flushright}
TIFR/TH/18-19
\end{flushright}

\begin{center}
\vspace*{1mm}

\vspace{1cm}
{\Large\bf 
Reconciling $\pmb{B}$-decay anomalies with neutrino masses, 
\vspace*{1mm}\\ 
dark matter and constraints from flavour violation
}

\vspace*{0.8cm}

{\bf Chandan Hati $^{a}$, 
Girish Kumar $^{b}$,  Jean Orloff $^{a}$ 
and  Ana M. Teixeira $^{a}$}  
\vspace*{.2cm}   

$^{a}$  Laboratoire de Physique de Clermont, CNRS/IN2P3 -- UMR 6533,\\ 
Campus des C\'ezeaux, 4 avenue Blaise Pascal, F-63178 Aubi\`ere Cedex, France
\vspace*{.2cm} 

$^{b}$ Department of Theoretical Physics, Tata Institute of
Fundamental Research, \\ 
Mumbai 400005, India
\end{center}

\vspace*{6mm}
\begin{abstract}
Motivated by an explanation of the $R_{K^{(*)}}$  anomalies, we
propose a Standard Model extension via two scalar SU(2)$_L$  triplet 
leptoquarks and three generations of triplet Majorana fermions.  
The gauge group is reinforced by a $Z_2$ symmetry, ensuring 
the stability of the lightest $Z_2$-odd particle, which is a
potentially viable dark matter candidate. Neutrino mass generation
occurs radiatively (at the three-loop level), and leads to important
constraints on the leptoquark couplings to leptons.  
We consider very generic textures for the flavour structure of the 
$h_1$ leptoquark Yukawa couplings, identifying 
classes of textures which succeed in saturating the $R_{K^{(*)}}$ anomalies. 
We subsequently carry a comprehensive analysis of the model's 
contributions to numerous high-intensity observables 
such as meson oscillations and decays, as well as charged lepton
flavour violating processes, which put severe constraints on the flavour
structure of these leptoquark
extensions. 
Our findings suggest that the most constraining observables 
are $K^+ \to \pi^+ \nu \bar \nu$ decays, 
and charged lepton flavour violating 
$\mu -e$ conversion in nuclei (among others). 
Nevertheless, for several classes of flavour textures 
and for wide mass regimes of the new mediators (within collider
reach), this Standard Model
extension successfully addresses neutrino mass generation, explains
the current $R_{K^{(*)}}$ tensions, and offers a viable dark matter
candidate.    
\end{abstract}

\newpage
\section{Introduction}

Despite the many successes of the Standard Model (SM)
in interpreting experimental data and
predicting new phenomena, three independent observations signal the 
need to consider New Physics (NP) scenarios: neutrino oscillations,
the baryon asymmetry of the Universe (BAU), and the lack of a
viable dark matter (DM) candidate. 
Although no direct evidence for NP states has been unveiled at the
LHC, certain
experimental measurements have revealed non-negligible tensions
with respect to the SM predictions. In addition to the anomalous
magnetic moment of the muon, recent data
hints to several discrepancies with respect to the SM 
in some $B$-meson decay modes, potentially 
suggesting the violation of lepton flavour universality (LFUV).
In the SM's original formulation, 
gauge interactions (both charged and neutral currents) 
are strictly lepton flavour universal; precise measurements of related
electroweak observables - for instance $Z\to \ell \ell$
decays~\cite{Patrignani:2016xqp} - have so far been in agreement
with the SM's predictions.  

The so-called $R_{K^{(\ast)}}$ observables are built from the 
comparison of the branching ratios (BR) of $B$
into di-muon and di-electron plus $K^{(*)}$ 
final states, and are parametrised as
\begin{equation}
R_{K^{(\ast)}} \,= \,
\frac{\text{BR}(B \to K^{(*)}\, \mu^+\,\mu^-)}{\text{BR}(B \to
  K^{(*)}\, e^+\,e^-)}\,.
\end{equation}
In the above ratio of BRs, the hadronic uncertainties cancel out 
to a very good approximation, and
consequently these observables are sensitive probes of 
NP contributions~\cite{Hiller:2003js}.  
First results on the measurement of $R_K$ by LHCb were reported in 
2014~\cite{Aaij:2014ora},
\begin{equation}
R_{K \,[1, 6]}^\text{LHCb}\,=\,0.745\pm_{0.074}^{0.090}\pm 0.036,
\end{equation}
having been obtained (as denoted in subscript) for the 
dilepton invariant mass squared bin $q^{2}\in [1,6]~\text{GeV}^{2}$. 
This corresponds to a $2.6 \sigma$ deviation from the SM prediction 
of $R_{K}^{\text{SM}}=1.00\pm0.01$~\cite{Bordone:2016gaq,Capdevila:2017bsm}. The
corresponding measurements for the decays into $K^* \ell \ell$ were
reported in 2017~\cite{Aaij:2017vbb},
\begin{equation}
R_{K^*[0.045, 1.1]} \,= \,0.66 ^{+0.11}_{-0.07} \pm 0.03\,,
\quad \quad
R_{K^* [1.1, 6]} \,= \,0.69 ^{+0.11}_{-0.07} \pm 0.05\,,
\end{equation}
respectively corresponding to $2.3\sigma$ and $2.6\sigma$ deviations 
from the expected SM values, 
$R_{K^*[0.045, 1.1]}^{\text{SM}} \sim 0.92\pm 0.02$ and 
$R_{K^* [1.1, 6]}^{\rm{SM}} \sim1.00\pm0.01$ \cite{Bordone:2016gaq,Capdevila:2017bsm}. 

Interestingly, deviations from SM expectations have also been
observed in $B \to D^{(*)} \ell \bar\nu$ decays, in particular in the
ratio of tau to final states composed of light leptons,
\begin{equation}
R_{D^{(*)}} \,= \,\frac{\Gamma(B \to D^{(*)} \,\tau^- \,\bar\nu)}{
\Gamma(B \to  D^{(*)}\, \ell^- \,\bar\nu)}\, \quad 
(\ell=e,\mu)\,.
\end{equation}
The measured value for $R_D =0.407 \pm 0.039\pm 0.024$ \cite{FLAG}, 
reported by several 
experiments~\cite{Lees:2012xj,Lees:2013uzd, Huschle:2015rga, Adachi:2009qg,
  Bozek:2010xy, Aaij:2015yra,Hirose:2016wfn}, 
already deviates from the SM prediction
$R_D^\text{SM}=0.299\pm 0.003$~\cite{Bigi:2016mdz}  
by about $2.3\,\sigma$. The experimental value of 
$R_{D^\ast}=0.304\pm
0.013\pm 0.007$~\cite{FLAG,Aaij:2015yra,Hirose:2016wfn,Amhis:2016xyh} 
is also larger than the SM expectation ($R_{D^\ast}^{\rm SM}=0.260\pm
0.008$~\cite{Bigi:2017jbd}), exhibiting a $2/6\sigma$
deviation. 
When combined, the latter experimental results point towards a 
deviation of 4.1\,$\sigma$ from the SM
prediction~\cite{FLAG,Ligeti:2016npd,Crivellin:2016ejn, Amhis:2016xyh}.
Other anomalies in $B$ meson decays have
emerged concerning the angular 
observable $P_5^{\prime}$ in $B \to K^\ast \ell^+ \ell^-$ processes. 
While LHCb's results for $P_5^{\prime}$ in $B \to K^\ast \mu^+
\mu^-$ decays manifest a slight discrepancy with respect to the SM 
(either due to NP contributions, or possibly a result of SM QCD
effects~\cite{Capdevila:2017ert}), the Belle 
Collaboration~\cite{Wehle:2016yoi} reported that when compared to the
muon case, $P_5^{\prime}$ results for electrons show a 
better agreement with respect to the SM prediction.
The $P_5^{\prime}$ results could thus be interpreted as suggestive of the
fact that NP effects may be dominant for the second generation of
leptons. 

In view of the above tensions,numerous well-motivated beyond the SM (BSM)
scenarios have been proposed in order to 
address one (or more) of these anomalies. Many NP models
in which LFUV effects arise at the loop level do not succeed in
explaining the $B$ meson anomalies; this has fuelled the interest to
consider BSM constructions capable of inducing LFUV
both at tree and loop level, as in the case of 
models with additional gauge bosons
($Z^\prime$), SM extensions via leptoquarks and other NP models
(see, for
example~\cite{Datta:2013kja,Ghosh:2014awa,Glashow:2014iga,Bhattacharya:2014wla,Freytsis:2015qca,Bardhan:2016uhr,Capdevila:2017bsm,Ghosh:2017ber,Ciuchini:2017mik,Choudhury:2017qyt,Choudhury:2017ijp}
for model independent
studies,~\cite{Altmannshofer:2014cfa,Crivellin:2015mga,Crivellin:2015lwa,Sierra:2015fma,Crivellin:2015era,Celis:2015ara,Bhatia:2017tgo,Kamenik:2017tnu,Camargo-Molina:2018cwu}
for $Z^\prime$
extensions,~\cite{Hiller:2014yaa,Gripaios:2014tna,Sahoo:2015wya,Varzielas:2015iva,Alonso:2015sja,Bauer:2015knc,Hati:2015awg,Fajfer:2015ycq,Das:2016vkr,Becirevic:2016yqi,Sahoo:2016pet,Cox:2016epl,Crivellin:2017zlb,Becirevic:2017jtw,Cai:2017wry,Buttazzo:2017ixm,Dorsner:2017ufx,Blanke:2018sro,Greljo:2018tuh,Bordone:2018nbg,Sahoo:2018ffv,Becirevic:2018afm}
for leptoquark models
and~\cite{Greljo:2015mma,Arnan:2017lxi,Geng:2017svp,Altmannshofer:2017poe,Das:2017kfo,Earl:2018snx}
for further examples).  

The leptoquark hypothesis - both in its scalar and vector field
realisations - has been extensively explored in recent years, as have
its implications for both quark and flavour dynamics.
In particular, having
leptoquark couplings which are necessarily flavour non-universal in
the lepton sector has multiple implications concerning lepton
observables, ranging from the anomalous moment of the muon, to 
charged lepton flavour violating (cLFV) decays and transitions. 
Likewise, efforts have been made to connect the 
neutrino mass generation problem (itself calling upon a modified lepton sector)
with an explanation of the flavour tensions via leptoquarks; many 
such models lead to radiative generation of the light
neutrino masses \footnote{For a recent review of radiative neutrino mass models see for example \cite{Cai:2017jrq}.}, for instance through mixings with the standard model
Higgs boson (in SM extensions via vector
leptoquarks~\cite{Deppisch:2016qqd}), using scalar leptoquarks and color-octet Majorana fermion~\cite{Cai:2017wry}, or calling upon a ``coloured
Zee-Babu model''~\cite{Zee:1985id,Babu:1988ki} (with the addition of 
scalar leptoquarks and diquarks), which leads to  
two-loop radiative neutrino mass generation~\cite{Guo:2017gxp}.  
Extensions of SM via both leptoquarks and additional Majorana
fermions aimed at connecting $B$-decay anomalies to neutrino masses
(radiatively generated at the three-loop level),
and to a solution of 
the dark matter problem~\cite{Nomura:2016ezz,Cheung:2016frv},
relying on the so-called 
``coloured KNT models''~\cite{Krauss:2002px}. These studies
also evaluated the impact of the BSM construction to cLFV decays,
focusing on the r\^ole of radiative charged lepton decays, 
$\ell \to \ell^\prime \gamma$.

Building upon the previous analysis, in this work we consider a scalar
leptoquark model, which aims at simultaneously explaining the $B$
meson decay anomalies, accounting for neutrino oscillation data and putting
forward a viable dark matter candidate. 
The SM is extended via two scalar
leptoquarks $h_{1,2}$ and three generations of triplet neutrinos $\Sigma_R^i$.  
The SM symmetry group is enlarged by a discrete $Z_2$ symmetry
under which only $h_2$ and $\Sigma_R^i$ are odd. 
While effectively forbidding the realisation of a tree-level 
type III seesaw~\cite{Foot:1988aq}, the $Z_2$
symmetry is instrumental to ensure the stability of the lightest  
$Z_2$-odd particle (the neutral component of the lightest triplet), 
which is found to be a viable dark matter candidate. 
Neutrino masses can be radiatively generated and, as
we argue here, complying with oscillation data turns out to severely
constrain the leptoquark Yukawa couplings. 
Focusing on saturating the $R_{K^{(*)}}$ anomalies, 
we carry a thorough analysis of this phenomenological model.
Our study relies in assuming generic perturbative textures for
the leptoquark Yukawa couplings: in particular, and 
contrary to previous analyses, we do not forbid couplings of the
leptoquarks to the first generation of quark and leptons. Moreover
we take into account a comprehensive set of flavour observables 
(meson and lepton rare decays and transitions); 
this allows to identify several
classes of textures for the leptoquark Yukawa couplings in agreement
with observation.
Our findings suggest
that the most severe constraints arise from $K \to \pi \nu \bar \nu$
decays and neutrinoless $\mu-e$ conversion in nuclei - and not from
radiative muon decays, as suggested by other studies; 
furthermore, the joint interplay of these high-intensity observables
also disfavours several ans\"atze for the leptoquark textures
previously considered (see~\cite{Nomura:2016ezz,Cheung:2016frv}).

The paper is organised as follows: after 
presenting the building blocks of the model 
in Section~\ref{sec:model}, and discussing  
neutrino mass generation 
in Section~\ref{sec:neutrino}, Section \ref{sec:dm} is devoted to
establishing first constraints on the model from the requirement of
having a viable DM candidate. 
The $B$ meson anomalies are presented in Section~\ref{sec:meson}, and 
the discrepancy between SM prediction and observation is parametrised in
terms of the leptoquark couplings. Sections~\ref{sec:mesoncon}
and~\ref{sec:lfv} are dedicated to the constraints potentially arising
from rare meson processes and from cLFV decays. 
Finally, our results (both in what concerns identifying viable
textures for the leptoquark Yukawa couplings, as well as numerical
studies of the model's parameter space) are collected in 
Section~\ref{sec:results}. 
We summarise the most important points, as well as our final remarks, 
in the Conclusions.

\section{SM extensions via scalar leptoquarks and Majorana 
triplets}\label{sec:model}

In this analysis we consider a SM extension in which two scalar
leptoquarks $h_{1,2}$ and three generations of
Majorana triplets $\Sigma_R^i$ are introduced, respectively with 
SU(3)$_C\times$SU(2)$_L\times$U(1)$_Y$ charge assignments of 
$(\bar{\textbf{3}}, \textbf{3},-1/3)$ and $(\textbf{1},
\textbf{3},0)$. 

As highlighted in the Introduction, the primary goals of this model
are to simultaneously address the problem of neutrino mass
generation, and provide a  viable DM candidate, while explaining the
observed anomalies in $B$ meson decays. 
If sufficiently long-lived or stable, the neutral component of
the lightest triplet
$\Sigma_R^1$ gives a potential cold dark matter (CDM) candidate: the
quantum corrections generate a mass splitting such that the neutral 
component is indeed the lightest one; its
stability can be ensured by reinforcing the SM gauge group by a
discrete $Z_2$ symmetry under which both $h_2$ and $\Sigma_R^i$ are odd,
while all the SM fields and $h_1$ are even. Since - and as mentioned
before - $\Sigma_R^1$ is the lightest state, the final DM
relic abundance is solely governed 
by the relevant electroweak (EW) gauge interactions and  $m_{\Sigma_R^1}$,
independent of its Yukawa interactions. This is in contrast with
scenarios in which a SU(2)$_L$ singlet fermion is considered as 
a dark matter candidate, subject only to Yukawa interactions.

It is important to notice that a consequence of the 
$Z_2$ symmetry is that it forbids a
conventional type III seesaw mechanism; however, neutrino
masses can still be radiatively generated at higher orders (as
discussed in the following section), from
diagrams involving the new exotic states and down-type quarks. 

The complete particle spectrum is presented in Table~\ref{table1}.  

{\small
\begin{table}[htb!]
\begin{center}
\begin{tabular}{|c|c|l|r|}
\hline
& Field	& {\small SU(3)$_C\times$SU(2)$_L\times$U(1)$_Y$ }& $Z_{2}$ \\
\hline \hline
Fermions & $Q_L \equiv(u, d)^T_L$&  \hspace*{10mm} $(\textbf{3},
\textbf{2},1/6)$ & $1$	\\
& $u_R$ & \hspace*{10mm} $(\textbf{3}, \textbf{1},2/3)$	& $1$	\\
& $d_R$ & \hspace*{10mm} $(\textbf{3}, \textbf{1},-1/3)$ & $1$	\\ 
& $\ell_L \equiv(\nu,~e)^T_L$ &\hspace*{10mm} $(\textbf{1},
\textbf{2}, -1/2)$	&  $1$	\\  
& $e_R$ & \hspace*{10mm} $(\textbf{1}, \textbf{1}, -1)$	&  $1$	\\ 
& $\Sigma_R$ & \hspace*{10mm} $(\textbf{1}, \textbf{3},  0)$ & $-1$\\ 
\hline
Scalars	& $H$ & \hspace*{10mm} $(\textbf{1}, \textbf{2},1/2)$	&   $1$	\\ 
& $h_1$	& \hspace*{10mm} $(\bar{\textbf{3}}, \textbf{3},  -1/3)$ &   $1$\\ 
& $h_2$	& \hspace*{10mm} $(\bar{\textbf{3}}, \textbf{3},  -1/3)$ &   $-1$ \\   
\hline \hline
\end{tabular}
\caption{Particle content and associated charges under the SM gauge
  group and additional discrete symmetries.}\label{table1}
\end{center}
\end{table}
}

The Lagrangian of the present SM extension can be cast as 
\begin{equation}
\mathcal{L} \, =\, \mathcal{L}^\text{SM}_\text{int} \, + \,
\mathcal{L}^{h,\Sigma}_\text{int} \, +\,
\mathcal{L}^{\Sigma}_\text{mass}\, - \,
V^{H,h}_\text{scalar}\,,
\end{equation}
in which $\mathcal{L}^{h,\Sigma}_\text{int}$ and 
$V^{H,h}_\text{scalar}$ respectively denote the interactions of 
$h_1$, $h_2$ and $\Sigma_R^i$ with matter\footnote{The
  couplings of $h_2$ to right-handed up-type quarks are absent
  due to hypercharge invariance, as pointed out in~\cite{Chen:2014ska}.}, 
and the scalar potential, while 
$\mathcal{L}^{\Sigma}_\text{mass}$ encodes the Majorana mass term for
the fermion triplets.
The new interaction and Majorana mass terms are given by 
\begin{align}\label{lag1}
\mathcal{L}^{h,\Sigma}_\text{int} \,  = \, 
y_{ij}\, \bar{Q}_{L}^{C\,i} \,\epsilon \,(\vec{\tau}. \vec{h}_{1})
\,L_{L}^{j}\,+\,z_{ij}\,\bar{Q}_{L}^{C\,i} \,\epsilon\,
(\vec{\tau}. \vec{h}_{1})^\dagger \,Q_{L}^{j}\,+\,
\tilde{y}_{ij}\,\overline{(\vec{\tau}. \vec{\Sigma})}^{C\,i,ab}_{R}
      [\epsilon \,(\vec{\tau}. \vec{h}_{2})\,\epsilon^{T}]^{ba}\,
      d_{R}^{j}
\, +\,\text{H.c.} \,,
\end{align}
\begin{align}\label{lag1.mass}
\mathcal{L}^{\Sigma}_\text{mass} \, =\,
-\frac{1}{2}\,\overline{\Sigma^{C}}^{i} \,M_{ij}^{\Sigma}\, \Sigma^{j}\,.
\end{align}
In the above $i,j=1\dots 3$ denote generation indices, 
while $a,b=1,2$ are SU(2) indices;  
$\tau^c$ are the Pauli matrices ($c=1,2,3$), and we have further
defined $ \epsilon^{ab}=(i \tau^2)^{ab}$.
Finally, $C$ denotes charge conjugation.
The scalar potential (including SM terms) can be written as 
\begin{align}\label{lag3.1}
V(H,h_1,h_2) \, = & \, 
\mu_H^2 H^\dag H \,+\,\frac{1}{2}\lambda_{H} \,|H^\dag H|^2 \, +\,  
\mu_{h_1}^2 \text{Tr}[h_1^{\dag } h_1] \,+\, \mu_{h_2}^2 \text{Tr}[h_2^{\dag } h_2] 
\,+ \nonumber\\
&+\, \frac{1}{8}\lambda_{h_1} \,[\text{Tr}(h_1^{\dag} h_1)]^2\,+
\, \frac{1}{8}\lambda_{h_2} \,[\text{Tr}(h_2^{\dag} h_2)]^2 \,+\,
\frac{1}{4}\lambda^{\prime}_{h_1} \,\text{Tr}[(h_1^{\dag }h_1)]^2 \,+\,
\frac{1}{4}\lambda^{\prime}_{h_2} \,\text{Tr}[(h_2^{\dag }h_2)]^2 \,+\nonumber\\
&+\, \frac{1}{2}\lambda_{H h_1}\, (H^\dag H) \,\text{Tr}[h_1^{\dag } h_1] 
\,+\, \frac{1}{2}\lambda'_{H h_1} \,\sum_{i=1}^3(H^\dag \,\tau_i \,H) 
\,\text{Tr}[h_1^{\dag } \,\tau_i\, h_1]\,+
\nonumber\\
&+ \, \frac{1}{2}\lambda_{H h_2}\, (H^\dag H) \,\text{Tr}[h_2^{\dag } h_2] 
\,+ \, \frac{1}{2}\lambda'_{H h_2} \,\sum_{i=1}^3(H^\dag \,\tau_i \,H) \,
\text{Tr}[h_2^{\dag } \,\tau_i \,h_2]\,+\nonumber\\
&+ \, \frac{1}{4}\lambda_{h} \,\text{Tr}[h_1^{\dagger}\, h_2 ]^2 + 
\,\frac{1}{8}\lambda'_{h} \,[\text{Tr}\,(h_1^{\dagger}\, h_2)]^2\,+\,
\frac{1}{4}\lambda''_{h} \,\text{Tr}[h_1^{\dagger}\, h_1] \,
\text{Tr}[h_2^{\dagger}\, h_2] \,+\,\text{H.c.} \,.
\end{align}

As can be inferred from inspection of Eq.~(\ref{lag1}), 
the simultaneous presence of the first two terms violates baryon
number, and can thus lead to $B-L$ conserving dimension-6 contributions
to proton decay. 
In the following analysis, we
will assume that these interactions are absent (i.e.,
$z_{ij}=0$), an hypothesis that can naturally arise from the embedding of
the model into an ultraviolet (UV) complete framework, as discussed 
in~\cite{Dorsner:2009cu,Hati:2018cqp}. 
The absence of the diquark couplings then allows to 
unambiguously assign baryon and lepton number to the scalar
leptoquarks $h_{1,2}$.

To cast the interaction Lagrangian in a more explicit way, it is
convenient to work in the U(1)$_\text{em}$ basis: the
physical states, respectively with electric charges 
$4/3, -2/3, 1/3$ can be written in terms of the SU(2) components as
follows
\begin{align}\label{eq:hsigma:redef}
& h^{4/3}_j\, =\, \frac{1}{\sqrt{2}} \left(h^{(1)}_j\,-\, i h^{(2)}_j\right)\,,
\quad 
h^{-2/3}_j\, =\, \frac{1}{\sqrt{2}} \left(h^{(1)}_j\,+\, i h^{(2)}_j\right)\,,
\quad  
h^{1/3}_j\, =\, h^{(3)}_j \quad (j=1,2)\,; \nonumber \\
& 
\Sigma^{+}\,=\, \frac{1}{\sqrt{2}} \left(\Sigma^{(1)}\,-\,i
\Sigma^{(2)}\right) \,,
\quad 
\Sigma^{-}\,=\, \frac{1}{\sqrt{2}} \left(\Sigma^{(1)}\,+\,i
\Sigma^{(2)}\right) \,,
\quad   
\Sigma^{0}\,=\, \Sigma^{(3)} \quad \text{\small(for the 3 generations)}\,.
\end{align}
Using the above redefinitions, the interaction
Lagrangian of Eq.~(\ref{lag1}) can be rewritten as 
\begin{align}
\mathcal{L}^{h,\Sigma}_\text{int} \,  = &\, 
- y_{ij}\, \bar{d}_{L}^{C\,i} \, h^{1/3}_{1}\, \nu_{L}^{j}
-\sqrt{2} \, y_{ij}\, \bar{d}_{L}^{C\,i} \, h^{4/3}_{1}\, e_{L}^{j}
+\sqrt{2} \, y_{ij}\, \bar{u}_{L}^{C\,i} \, h^{-2/3}_{1}\, \nu_{L}^{j}
-y_{ij}\, \bar{u}_{L}^{C\,i} \, h^{1/3}_{1} \, e_{L}^{j}
\nonumber\\ 
& -2 \, \tilde{y}_{ij}\, \overline{\Sigma^{0}}^{C\,i}_{R} \,  h^{1/3}_{2}
\, d_{R}^{j}
-2 \, \tilde{y}_{ij}\, \overline{\Sigma^{+}}^{C\,i}_{R}\, h^{-2/3}_{2} \, d_{R}^{j}
-2\, \tilde{y}_{ij}\, \overline{\Sigma^{-}}^{C\,i}_{R} \,  h^{4/3}_{2}\,  d_{R}^{j}
+ \text{H.c.}\; . 
{\label{lag2}}
\end{align}
Following the decomposition of
Eq.~(\ref{eq:hsigma:redef}), the most relevant interaction term for
neutrino mass diagrams can now be written as 
\begin{equation}
\frac{\lambda_{h}}{4}\,
\text{Tr}(h_1^{\dagger} \,h_2 \,h_1^{\dagger} \,h_2)\, =\,
\frac{\lambda_{h}}{2}\,
h_{1}^{-1/3}\,h_{2}^{1/3}\,h_{1}^{-1/3}\,h_{2}^{1/3}\,-\,
\lambda_{h}\,
h_{1}^{-1/3}\,h_{2}^{-2/3}\,h_{1}^{-1/3}\,h_{2}^{4/3}\,.
\end{equation}

In what follows (and for simplicity), we will further assume 
the couplings $\lambda^\prime_{Hh_{1,2}}$ to be negligible; this leads
to having degenerate physical masses for the components of the
scalar triplets $h_{1,2}$ 
($m^2_{h_{1(2)}}=\mu_{h_{1(2)}}^2+\lambda_{Hh_{1(2)}}v^2/2$, with $v$
the SM Higgs vacuum expectation value), and 
thus allows to comply with EW precision 
constraints on oblique parameters.

\section{Radiative neutrino mass generation and leptonic 
mixings}{\label{sec:neutrino}}

As mentioned in the previous section, the $Z_2$ symmetry
precludes any coupling between the fermion triplets and neutral
leptons, which effectively dismisses a type III
seesaw explanation of 
neutrino mass generation (at the tree level). The absence
of right-handed neutrinos further excludes the possibility of
Dirac-type masses for the neutral leptons. 
Interestingly, the particle content of the model does allow a natural 
explanation to the smallness of neutrino masses: these are radiatively
generated, from  higher order contributions.    

\begin{figure}[h!]
\begin{center}
\includegraphics[width=0.49\textwidth]{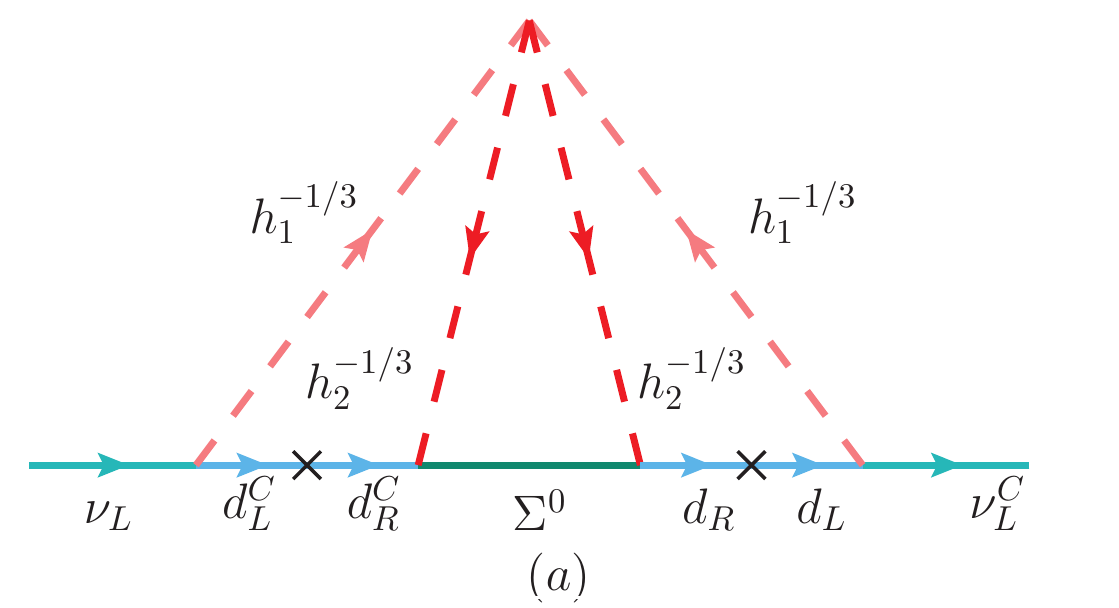}
\includegraphics[width=0.49\textwidth]{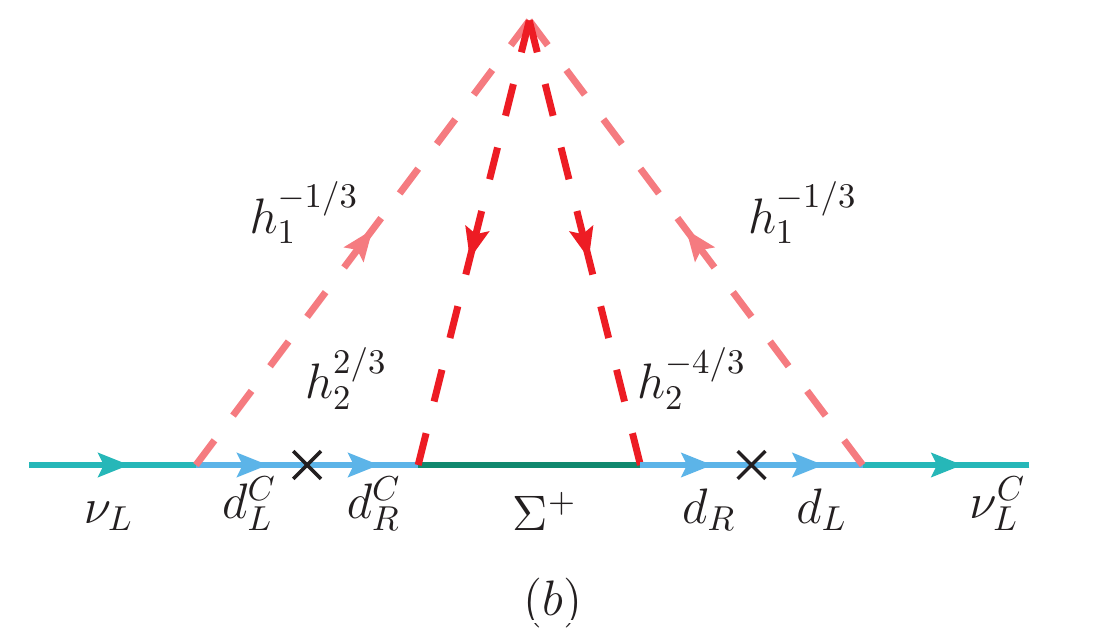}
\caption{Three-loop diagrams contributing to neutrino masses, mediated
  via neutral (charged) fermion triplets.
  A third diagram (not displayed here) can be obtained by replacing
  $\Sigma^{+}$ by its charge conjugate state $\Sigma^{-}$, while also
  exchanging the internal $h_2^{2/3, -4/3}$ propagators of panel (b).}  
\label{fig:numass}
\end{center}
\end{figure}

The first non-vanishing contributions to $m_\nu$ arise at
the three-loop level, and are induced by the diagrams 
displayed in
Fig.~\ref{fig:numass}, from the exchange of leptoquarks $h_{1,2}$ and
neutral (charged) fermion triplets, calling upon chirality flips in
the internal down-type quark lines (proportional to the down quark masses). 
Despite the different particle content, the diagrams 
are akin to those originally proposed
in~\cite{Krauss:2002px}, which were mediated via 
colourless scalars and a $Z_2$-odd singlet Majorana fermion.  
(In the latter case, the computation of the contributions to $m_\nu$
has been carried in~\cite{Ahriche:2013zwa}.)

The computation of the different diagrams of Fig.~\ref{fig:numass}
leads to the following contributions to neutrino masses (which are for
simplicity cast in the weak interaction basis): 
\begin{align}
-i (m_{\nu})_{\alpha\beta}  \, =\, &
2\int \frac{d^{4}Q_{1}}{\left(2\pi\right)^{4}}\,
\int \frac{d^{4}Q_{2}}{\left(2\pi\right)^{4}}\,
\frac{i}{Q_{1}^{2}\,-\,m_{h_{1}}^{2}}\,
\left(-2i\,y^{T}_{\alpha i}\right)\,  
P_{L}\, \frac{i}{\slashed {Q}_{1}\,-\,m_{D_{i}}}\, \times \nonumber \\
&
\int \frac{d^{4}k}{\left(2\pi\right)^{4}}\, 
\left(-i\,\tilde{y}^{T}_{ij}\right)\, P_{L} \,
\frac{i\left(\slashed k \, +\,m_{\Sigma_{j}}\right)}{k^{2}-m_{\Sigma_{j}}^{2}}\, 
P_{L}\,\left(-i\, \tilde{y}_{jk}\right)\,
\frac{i}{\left(k+Q_{1}\right)^{2}\,-\,m_{h_{2}}^{2}} 
\times \nonumber \\
&
\left[-i(s_a \times \kappa_c)\, \frac{\lambda_{h}}{2} -2i (s_b \times 
\kappa_c) \, \lambda_{h} \right]\,
\times\nonumber\\
&
\frac{i}{\left(k\,-\,Q_{2}\right)^{2}\,-\, m_{h_{2}}^{2}}\,
\frac{i}{\slashed Q_{2}\,-\,m_{D_{k}}}\, P_{L}\, 
\left(-2i\, y_{k \beta}\right)\, \frac{i}{Q_{2}^{2}\,-\,m_{h_{1}}^{2}},
\end{align}
in which $Q_{1,2}$ denotes the momenta of the internal down-type
quarks, $m_D =\text{diag}(m_d,m_s,m_b)$ 
is the diagonal down-quark mass 
matrix\footnote{For convenience, and without loss of generality, we
  chose to work in a basis in which the down-quark Yukawa couplings
  are taken to be diagonal ($Y^d_{ij} \delta_{ij} = m_{D_{i}}/v$),
  while parametrising the up-quark Yukawa couplings as 
  $Y^u_{ij} = V^\dagger_{ij} m_{U_{j}}/v$, with $V$ the
  Cabibbo-Kobayashi-Maskawa (CKM) mixing matrix.}, 
and $P_{L,R}=(1\mp\gamma_{5})/2$.
For clarity, we have explicitly highlighted the contributions of the 
four-scalar leptoquark vertices of diagrams (a) and (b) (as well as
the $\Sigma^{-}$ counterpart of (b)), writing them as products of 
symmetry and colour factors, respectively $s$ and $\kappa_c$. Diagram
(a) is associated with $s_a=4$, while (b)-like diagrams lead to
$s_b=2$; for the colour factor one has $\kappa_c=15$, due to 
the possible distinct contractions of the four-scalar leptoquark 
vertex (in agreement with~\cite{Cheung:2016frv}).

Using the appropriate loop 
integrals\footnote{In particular, one makes  use of the identity
\begin{equation}
\int \frac{d^{4}Q}{\left(2\pi\right)^{4}} 
\frac{1}{\left(Q^{2}-m_{0}^{2}\right)  
\left( Q^{2}-m_{1}^{2}\right)  
\left(\left(  k+Q\right)^{2}-m_{2}^{2}\right)}\,=\,i\,
\frac{B_{0}\left(k^{2},m_{1}^{2},m_{2}^{2}\right)\,-\,
B_{0}\left(k^{2},m_{0}^{2},m_{2}^{2}\right)}{16\pi^{2}\,m_{1}^{2}},
\end{equation}
where $B_{0}$ is the Passarino-Veltman function defined
(in terms of the renormalisation scale $\mu$) 
as~\cite{Passarino:1978jh} 
\begin{equation}
B_{0}\left(k^{2},m_{1}^{2},m_{2}^{2}\right)  \,=\, 
\frac{1}{\epsilon}-
\int_{0}^{1} dx \ln \left(
\frac{-x\,\left(1-x\right)\,k^{2}\,+\,\left(1-x\right)\,m_{1}^{2}
\,+\,x\,m_{2}^{2}}{\mu^{2}}\right)\,.
\end{equation}
},
and after a Wick rotation we obtain 
\begin{equation}{\label{eq:numass}}
(m_{\nu})_{\alpha\beta}\, =\, -30\, 
\frac{\lambda_{h}}{\left(4\pi^{2}\right)^{3}\, m_{h_{2}}}\, y^{T}_{\alpha i}\, 
m_{D_{i}} \, \tilde{y}^{T}_{ij} \, 
G\left(\frac{m_{\Sigma_{j}}^{2}}{m_{h_{2}}^{2}},
\frac{m_{h_{1}}^{2}}{m_{h_{2}}^{2}}\right)\, 
\tilde{y}_{jk}\, m_{D_{k}}\, y_{k \beta}   ,
\end{equation}
where $m_D$ is again the diagonal down type mass matrix 
(in the computation of the loop integrals, the down-quark masses are
neglected when compared to the heavier $h_{1,2}$ and $\Sigma$ masses
in the loop), and we recall that $y$ ($\tilde y$) denotes the Yukawa
couplings of the $Z_2$-even leptoquark $h_1$ to matter ($Z_2$-odd
leptoquark $h_2$ and triplet fermion $\Sigma_R$ to down quarks);  
finally $G(a,b)$ is defined as 
\begin{equation}\label{FF}
G\left(a,b\right) \, = \,\frac{\sqrt{a}}{8\,b^{2}}\,
\int_{0}^{\infty} dr\,\frac{r}{r+a}\,\left[
\int_{0}^{1} dx \ln  \left(\frac{x\,
\left(1-x\right)\, r \,+\, \left(1-x\right)\,b\,+\,x}{x \,
\left(1-x\right)\,r\,+\,x}\right)\right]^{2}\,, 
\end{equation}
and in the case in which (for simplicity and without loss of generality)
$m_\Sigma$ is assumed to be diagonal, 
$G({m_{\Sigma_{j}}^{2}}/{m_{h_{2}}^{2}},
{m_{h_{1}}^{2}}/{m_{h_{2}}^{2}})$ will also be diagonal. 

The neutrino mass eigenstates can be obtained using the transformation 
\begin{equation}\label{eq:numassdiag}
m_{\nu}^{\text{diag}}\,\equiv\, 
{\text{diag}}(m_{\nu_1},m_{\nu_2},m_{\nu_3})\,=\,
U^{T}_{i\alpha}\, (m_{\nu})_{\alpha \beta} \,U_{\beta i}\,, 
\end{equation}
where $U$ is the Pontecorvo-Maki-Nakagawa-Sakata (PMNS) 
unitary mixing matrix,  which we parametrise as follows
\begin{equation}
U\,=\,\left(
\begin{array}
[c]{ccc}%
c_{12}c_{13} & c_{13}s_{12} & s_{13}e^{-i\delta_{D}}\\
-c_{23}s_{12}-c_{12}s_{13}s_{23}e^{i\delta_{D}} & c_{12}c_{23}-s_{12}%
s_{13}s_{23}e^{i\delta_{D}} & c_{13}s_{23}\\
s_{12}s_{23}-c_{12}c_{23}s_{13}e^{i\delta_{D}} & -c_{12}s_{23}-c_{23}%
s_{12}s_{13}e^{i\delta_{D}} & c_{13}c_{23}%
\end{array}
\right) \cdot {\text{diag}}(1,e^{i\alpha/2},e^{i\beta/2})\,.
\end{equation}
In the above, $s_{ij}\equiv\sin(\theta_{ij})$ and 
$c_{ij}\equiv\cos(\theta_{ij}) $; $\delta_{D}$ is the CP violating 
Dirac phase while $\alpha$ and $\beta$ are Majorana phases.

It is important to notice that as can be seen from
Eqs.~(\ref{eq:numass}) and~(\ref{eq:numassdiag}), neutrino masses (and
leptonic mixings) do indeed depend on both of the Yukawa couplings
involving the leptoquarks,  
$y$ and $\tilde{y}$. In particular, the former will be at the source
of a number of flavour transitions, including rare meson decays,
neutral meson-antimeson oscillations as well as charged lepton flavour
violating processes. This implies that a strong connection between
neutrino phenomena and flavour nonuniversal
processes is established via the flavour structure of the Yukawa
matrix $y$.

In the absence of a complete framework proving a full theory of
flavour (which would suggest a structure for 
$y$ and $\tilde y$), one can nevertheless parametrise one of
the Yukawa couplings - for example, $\tilde y$ - using a modified  
Casas-Ibarra parametrisation~\cite{Casas:2001sr}, which further allows
to accommodate neutrino oscillation data. 

In order to construct the modified Casas-Ibarra parametrisation,  
we first notice that from Eqs.~(\ref{eq:numass})
and~(\ref{eq:numassdiag}) one can write the diagonal neutrino mass
matrix as
\begin{align}
m^\text{diag}_\nu \,= \, 
U^T\; y^T\; m_D\; \tilde{y}^{T}\; F(\lambda_h, m_\Sigma, m_{h_{1,2}})
\; \tilde{y}\; m_D\; y\; U\,,
\label{eq:cip1}
\end{align}
in which we have omitted flavour (generation) indices for simplicity, and 
where
\begin{align}
F(\lambda_h, m_\Sigma, m_{h_{1,2}})\,=\,
\frac{30\,\lambda_h}{(4\pi^2)^3\, m_{h_2}} \,
G_j\left(\frac{m_{\Sigma_{j}}^{2}}{m_{h_{2}}^{2}},
\frac{m_{h_{1}}^{2}}{m_{h_{2}}^{2}}\right)\,.
\end{align}
As noted before, for a diagonal $m_\Sigma$, $G$ and thus $F$ are also
diagonal matrices in generation space, which allows to write the identity
\begin{align}
(\sqrt{m^\text{diag}_\nu}^{-1} U^T\; y^T\; m_D\; 
{\tilde y}^{T}\;\sqrt{F})\,
(\sqrt{F}\; \tilde y\; m_D\; y\;
U\;\sqrt{m^\text{diag}_\nu}^{-1})\,=\,
\mathbb{1}\,=\,{\cal R}^{T}\,{\cal R}, 
\label{eq:cip2}
\end{align}
with ${\cal R}$ an arbitrary complex orthogonal matrix (${\cal
  R}^{T}{\cal R}=\mathbb{1}$) which can be parametrised  
in terms of three complex angles, $\theta_i$ ($i=1-3$). 
Finally, from Eq.~(\ref{eq:cip2}) one obtains the modified  Casas-Ibarra
parametrisation, which allows to write the Yukawa couplings of the
$h_2$ leptoquark in terms of observable quantities (light neutrino
masses, leptonic mixings, down quark masses, triplet and leptoquark
masses), and of two unknown quantities - the $h_1$ leptoquark Yukawa
couplings and a complex orthogonal matrix - as  
\begin{align}
\tilde{y} \,= \, F^{-1/2}\; \mathcal {R}\; \sqrt{m^\text{diag}_\nu} 
\;U^\dag y^{-1} m_d^{-1}\,.
\label{eq:cip}
\end{align}
The above parametrisation (which is in agreement to a
similar approach carried in~\cite{Cheung:2016frv}), allows to write
the couplings $\tilde{y}$ in terms of $y$ up to a complex orthogonal
matrix.
As will be discussed in detail in Section~\ref{sec:texture}, once the
approximate flavour texture of $y$ is inferred from various
experimental constraints, that of $\tilde{y}$ can be derived (up to
the mixings due to $\mathcal{R}$).

\section{A viable dark matter candidate}\label{sec:dm}

Reinforcing the SM gauge group via a discrete $Z_2$ symmetry ensures
the stability of the lightest state which is odd under $Z_2$. If the
lightest $Z_2$-odd particle (LZoP) is neutral, and the strength of its 
interactions is such that its relic abundance is in agreement
with observational data, then it can be indeed a viable dark matter
candidate. 

In our analysis we assume that the spectrum of the new states is such
that one has $m_{\Sigma_R^1} < m_{h_2}$. 
At the tree level, all the components of a generation 
$\Sigma_R^i$ have the same mass
$m_{\Sigma^{i,\text{tree}}} =m_{\Sigma^{i,\pm}}=m_{\Sigma^{i,0}}$. 
The degeneracy between the components is broken by EW
radiative corrections, which render the charged states heavier than
the neutral one. Dropping for simplicity the generation indices (in
this section our discussion is focused on the components of the 
lighest triplet, $\Sigma_R^1$), 
the splitting between the neutral and charged
components is given by~\cite{Cirelli:2005uq} 
\begin{equation}
\Delta_{m_{\Sigma}}\,=\,m_{\Sigma^{\pm}}\,-\,m_{\Sigma^{0}}\,=\,
\frac{\alpha_2 \,m^\text{tree}_\Sigma}{4\pi}\,
\left[ f_\text{EW}\left(x_{W\Sigma} \right)\,-\,\cos^{2} \theta_w
  f_\text{EW}\left(x_{Z\Sigma}\right)\right]\,,
\end{equation}
with $x_{ij}=\frac{m_i}{m_j}$ and
\begin{eqnarray}
f_\text{EW}(x)\,=\,-x^2 \,+\,x^4 \,\ln x \,+\,
x\,(x^2-4)^{1/2}\, \left(1+\frac{x^2}{2}\right) 
\,\ln \left( -1- \frac{x}{2}\,(x^2-4)^{1/2}+\frac{x^2}{2}\right) \frac{x}{2},
\end{eqnarray}
and in which $\alpha_2=\alpha_{e}/\sin^{2} \theta_{w}$, with
$\alpha_{e}$ the fine structure constant and $\theta_w$
the weak mixing angle. In the limit $m_\Sigma\gg M_{W,Z}$ (justified
by negative searches at LHC and EW precision measurements),
the EW radiative corrections are found to be of order 
$m_{\Sigma^{\pm}}-m_{\Sigma^{0}}\sim 166$~MeV. While the latter mass
difference is enough to ensure that the neutral component of the
lightest triplet, $\Sigma^0$, is indeed the LZoP,
it is sufficiently small to be neglected in the subsequent
(numerical) analysis.

The relic abundance of the LZoP $\Sigma^{0}$ is determined by its 
interactions and by the annihilation and coannihilation channels open
in view of the particle mass spectrum. 
In addition to the Yukawa interactions with $h_2$, the $\Sigma_R$
triplets are subject to SU(2)$_L$ gauge interactions. Since, as
previously highlighted, we assume that $m_{\Sigma_R^1} < m_{h_2}$, the relic
density of $\Sigma^{0}$ is, to first order approximation\footnote{A
  full evaluation of the relic density would further call upon higher
  order effects, which would in turn depend on the Yukawa couplings of
  the LZoP, and other additional interactions. Such a study lies beyond the
  scope of the present analysis; here our primary goal is to identify 
  a viable dark matter candidate, and obtain
  indicative constraints on its mass.}, solely
determined by its gauge interactions, which govern the distinct
annihilation and coannihilation channels (involving also the charged
components $\Sigma^\pm$). 
In particular, the annihilation and coannihilation processes involve
the following channels: 
$\Sigma^{0}\Sigma^{0}\rightarrow W^{\pm}W^{\mp}$ 
through $t$-channel $\Sigma^{\pm}$ exchange; 
$\Sigma^{0}\Sigma^{\pm} \rightarrow W^{\pm} Z^{0}$  
via $t$-channel $\Sigma^{\pm}$ exchange; 
$\Sigma^{0}\Sigma^{\pm} \rightarrow W^{\pm}Z^{0},
W^{\pm}H, \bar{f}f^{\prime}$ (via $s$-channel $W^{\pm}$ exchange). 
Other processes involving the charged components of the triplet
fermion must be also taken into account in the Boltzmann equations  
leading to the computation of the relic density. These include 
$\Sigma^{\pm}\Sigma^{\mp}\rightarrow Z^{0}Z^{0} (W^{\pm}W^{\mp})$ 
through $\Sigma^{\pm(0)}$ $t$-channel exchange,
$\Sigma^{\pm}\Sigma^{\mp} \rightarrow W^{\pm}W^{\mp}, Z^{0}H,
\bar{f}f$ ($s$-channel via $Z^0$ mediation), and finally
$\Sigma^{\pm}\Sigma^{\pm} \rightarrow W^{\pm} W^{\pm}$ 
through $\Sigma^{0}$ exchange ($t$-channel).

The relevant cross sections for the above mentioned processes 
are given by~\cite{Ma:2008cu}
\begin{align}
& \sigma(\Sigma^{0}\Sigma^{0})|\bar v|\,\simeq \,
2\pi\,\frac{\alpha_2^2}{m_\Sigma^2}, \quad
\sigma(\Sigma^{0}\Sigma^{\pm})|\bar v|\,\simeq \,
2\pi\,\frac{29 \,\alpha_2^2}{16 \,m_\Sigma^2}\,, \nonumber\\
&\sigma(\Sigma^{+}\Sigma^{-})|\bar v|\,\simeq\, 
2\pi\,\frac{37 \,\alpha_2^2}{16 \,m_\Sigma^2}\,, \quad
\sigma(\Sigma^{\pm}\Sigma^{\pm})|\bar v|\,\simeq\, 
2\pi\,\frac{\alpha_2^2}{2 m_\Sigma^2}\,,
\end{align}
where only the coefficients $a_{ij}$ are kept in the relative velocity
($\bar v$)
expansion of the cross section, i.e. $\sigma_{ij}|\bar v|=
a_{ij}+b_{ij} \bar v^{2}$.

The computation of the relic abundance follows closely the method
of~\cite{Griest:1990kh}, in which the freeze-out temperature
of the LZoP $\Sigma^{0}$ ($x_{f}\equiv m_{\Sigma}/T_f$) 
is obtained in terms of the thermally averaged effective 
cross section $\langle \sigma_{\text{eff}} |\bar v|\rangle$.
The relevant channels above referred to contribute to the 
thermally averaged effective cross section $\langle
\sigma_{\text{eff}} |\bar v|\rangle$ as
\begin{eqnarray}
\langle \sigma_{\text{eff}}|\bar v|\rangle\,=\,&\frac{g_{0}^{2}}{g_{\text{eff}}^2}
\,\sigma(\Sigma^{0}\Sigma^{0})|\bar v| \,+\, 4\, 
\frac{g_{0}\, g_{\pm}}{g_{\text{eff}}^2}\, \sigma(\Sigma^{0}\Sigma^{\pm})|\bar v|
\,\left(1+\frac{\Delta_{m_{\Sigma}}}{m_{\Sigma}}\right)^{3/2} \exp 
\left(-\frac{\Delta_{m_{\Sigma}}}{m_{\Sigma}} \,x_f \right) \,+\nonumber\\
&+ \frac{g_{\pm}^2}{g_{\text{eff}}^2} \,
\left[\,2\,\sigma(\Sigma^{+}\Sigma^{-})|\bar v| \,+ \,
2\,\sigma(\Sigma^{\pm}\Sigma^{\pm})|\bar v| \right] 
\left(1+\frac{\Delta_{m_{\Sigma}}}{m_{\Sigma}}\right)^{3} \exp 
\left( -2\frac{\Delta_{m_{\Sigma}}}{m_{\Sigma}} x_f \right)\,,
\end{eqnarray}  
and the freeze-out temperature is then recursively given by
\begin{eqnarray}
x_{f}\,\equiv\, \frac{m_{\Sigma}}{T_f}\,=\,
\ln \left(
\frac{0.038 \; g_{\text{eff}} \; M_{\text{Pl}}\; m_{\Sigma}\; \langle
  \sigma_{\text{eff}} |\bar v|\rangle}{g_{\ast}^{1/2}
  x_f^{1/2}}\right) \,, 
\end{eqnarray}
in which
$g_{\ast}\sim 106.75$ is the total number of effective relativistic
degrees of freedom the freeze-out, and $M_{\text{Pl}}=1.22 
\times 10^{19}$~GeV is the Planck scale. 
The quantity $g_{\text{eff}}$ is related to the
degrees of freedom of the triplet
components,  $g_{0}=2$ and $g_{\pm}=2$, respectively for 
$\Sigma^{0}$ and $\Sigma^{\pm}$ and to $\Delta_{m_{\Sigma}}$
(the mass splitting between the 
charged and neutral components of $\Sigma_R$), and can be written as 
\begin{eqnarray}
g_{\text{eff}}\, = \,
g_{0}\,+\,2\,g_{\pm}\left(1+\frac{\Delta_{m_{\Sigma}}}{m_{\Sigma}}\right)^{3/2} 
\exp \left( -\frac{\Delta_{m_{\Sigma}}}{m_{\Sigma}}\, x_f \right)\,.
\end{eqnarray}
Finally, the relic abundance is given by
\begin{equation}\label{eq:relic}
\Omega \,h^{2}\,=\,
\frac{1.07 \times 10^9 \,x_f }{g_{\ast}^{1/2}\; M_{\text{Pl}}({\text{GeV}})\; I_a},
\end{equation}
with $I_a$ the annihilation integral, which is defined as
\begin{equation}
I_a\,=\,x_f \,\int_{x_f}^{\infty} x^{-2} \,a_{\text{eff}}\; dx\,.
\end{equation}
In the above, one has used the approximation
$a_{\text{eff}} \sim \sigma_{\text{eff}}|\bar v|$ (we recall that 
we have not taken into account the second and higher order terms in the
relative velocity expansion of the effective cross section). 

\begin{figure}[t!]
\begin{center}
\includegraphics[width=0.70\textwidth]{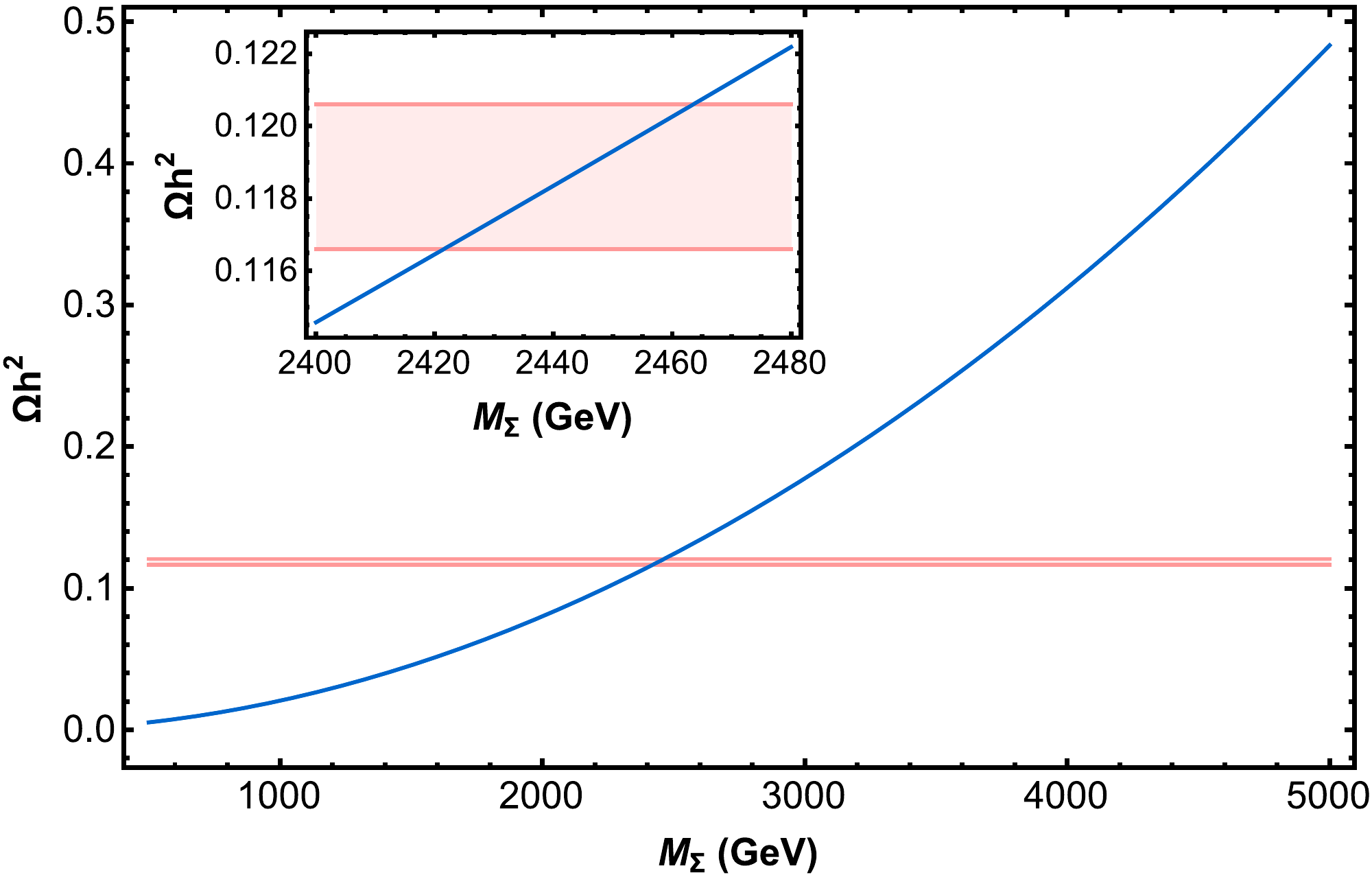}
\caption{Estimated relic abundance (see text for details) of the dark
  matter candidate, $\Sigma^{0}$, as a function of its mass. The
  rose-coloured band corresponds to the latest observational data 
  $\Omega h^{2}= 0.1198 \pm 0.0015$ \cite{Ade:2015xua}. 
 } 
\label{fig:dm}
\end{center}
\end{figure}

For our purpose of constraining the parameter
space of the model, we will rely on a simple
iterative solution of Eq.~(\ref{eq:relic}), which allows to 
infer first limits on the values of $m_{\Sigma^1_R}$ leading to
a relic density in agreement with the most recent observational 
data~\cite{Ade:2015xua}, 
\begin{equation}\label{eq:omegah2exp}
\Omega\, h^{2}\,=\, 0.1186 \,\pm \,0.0020\,.
\end{equation}
The results of this (approximative) numerical analysis are displayed
in Fig. \ref{fig:dm}, which reveals that having an LZoP mass in the
range $2.425\;{\text{TeV}}< m_{\Sigma}<2.465\;{\text{TeV}}$ 
leads to a dark matter relic abundance in agreement with 
the latest data. In what follows, we will use 
$m_{\Sigma}\sim 2.45 {\text{TeV}}$ 
as an illustrative benchmark value for the 
lightest fermion triplet mass.

Although already mentioned, we nevertheless stress again that 
several approximations were done in our computation of the 
relic abundance; moreover, one should also explicitly
solve the coupled Boltzmann equations to numerically obtain the
allowed mass range of the LZoP (the lightest $\Sigma^{0}$).
Other than complying with the dark matter relic abundance, the
potential candidate is also subject to the increasingly strong
constraints from direct and indirect searches 
(see, e.g.~\cite{Cirelli:2005uq,Choubey:2017yyn}); a detailed discussion 
of the latter (and the dedicated facilities) lies beyond
the scope of this work.

\section{Addressing the $\pmb{B}$ meson decay
  anomalies}\label{sec:meson}

As can be seen from the interaction Lagrangian of Eq.~(\ref{lag2}), 
the scalar leptoquark $h_1$ couples to both down-type quarks and
charged leptons, the couplings having a priori a non-trivial structure
in flavour space. This will open the door to new contributions to
numerous rare flavour changing transitions and decays, and - most
importantly - can potentially lead to lepton flavour non-universal
effects, such as those currently suggested by the reported LHCb
anomalies.

In this section we will explore to which extent the current model
succeeds in addressing the $B$ meson decay anomalies, $R_{K^{(\ast)}}$
and $R_{D^{(\ast)}}$, respectively associated with neutral and
charged current transitions. The abundant constraints arising from 
negative searches for NP effects in meson oscillation and decays, as
well as in charged lepton flavour violating observables will be discussed
in Sections~\ref{sec:mesoncon} and~\ref{sec:lfv}.

\subsection{Neutral current anomalies: $R_K$ and $R_{K^\ast}$}\label{subsec:RK}
As mentioned in the Introduction, recent 
measurements~\cite{Aaij:2014ora,Aaij:2017vbb} of the ratios of
branching ratios of $B\to K(K^\ast)\ell\ell$ decays into 
pairs of muons over those into di-electrons exhibit non-negligible
deviations when compared to the SM predictions~\cite{Bordone:2016gaq,Capdevila:2017bsm}; 
as already stated one has
\begin{eqnarray}
& R_{K [1,6]}\, =\, 0.745\,\pm_{0.074}^{0.090}\,\pm\,
0.036,\, \quad
& R_{K}^{\text{SM}}\, =\, 1.00, \pm\, 0.01 
 \nonumber\\
& R_{K^*[0.045, 1.1]}\, =\, 0.66 ^{+0.11}_{-0.07} \,\pm\,
 0.03\,, \quad  
&R_{K^*[0.045, 1.1]}^{\text{SM}}\,  \sim\,  0.92\pm 0.02 \nonumber\\
& R_{K^* [1.1, 6]} \,= \,0.69 ^{+0.11}_{-0.07}\, \pm
0.05\,, \quad
& R_{K^* [1.1, 6]}^{\text{SM}} \,\sim \, 1.00\pm 0.01\, ,
\end{eqnarray}
where the dilepton invariant mass squared bin (in $\text{GeV}^{2}$) 
is identified by the subscript. The comparison of SM predictions with
observation respectively reveals deviations of $2.6 \sigma$, $2.4\sigma$
and $2.5 \sigma$. 

For the leptoquark mass regime considered here (multi-TeV), 
the neutral current effects induced by the heavy degrees of freedom (SM
and NP contributions) in the quark level transitions $d_j \to d_i \ell^+
\ell^-$ can be described by the following effective 
Hamiltonian~\cite{Altmannshofer:2008dz,Becirevic:2016oho} 
\begin{eqnarray}\label{eq:effHbtos} 
\mathcal{H} (d_j \to d_i \ell^+\ell^-)
&= & -\frac{4\,G_F}{\sqrt{2}} V_{tj}\, V_{ti}^\ast
\left[ 
C_7^{ij}\, \mathcal{O}^{ij}_7 + C_{7^\prime}^{ij}\,\mathcal{O}^{ij}_{7^\prime}
\,+\hspace*{-4mm}\sum_{_{X=9,10,S,P}}\hspace*{-2mm}
\left(C_X^{ij;\ell\ell'}\, \mathcal{O}_X^{ij;\ell\ell'} \,+\, 
C_{X^\prime}^{ij;\ell\ell'}
\,\mathcal{O}_{X^\prime}^{ij;\ell\ell'}\right) \,+
\right.\nonumber\\    && 
\left. \phantom{[  
C_7^{ij}\, \mathcal{O}^{ij}_7 + C_{7^\prime}^{ij}\,\mathcal{O}^{ij}_{7^\prime}
\,+}
+\, C^{ij;\ell\ell'}_T\,\mathcal{O}^{ij;\ell\ell'}_T +\, C_{T5}^{ij;\ell\ell'}
\,\mathcal{O}^{ij;\ell\ell'}_{T5}\right] + \,\text{H.c.}\;,
\end{eqnarray}
in which $G_F$ is the Fermi constant and $V$ is the CKM mixing matrix.
The effective operators present in the above equation can be 
defined as (for simplicity, 
hereafter we drop the $ij$ superscripts, 
which are set to $i,j=s,b$ for the process $b \to s \ell^+ \ell^-$):
\begin{eqnarray}\label{eq:ops}
&&  \mathcal{O}_7 \,= \,\frac{e m_{d_j}}{(4\pi)^2}\, (\bar d_i
  \sigma_{\mu\nu} P_R d_j) 
  \,F^{\mu\nu}\,, \nonumber \\
&& \mathcal{O}_9^{\ell\ell'} \,= \,\frac{e^2}{(4\pi)^2}\, (\bar d_i
\gamma^\mu P_L d_j) 
  (\bar\ell \gamma_\mu \ell')\,, \quad
\mathcal{O}_{10}^{\ell\ell'} \,= \,\frac{e^2}{(4\pi)^2}\, (\bar d_i
\gamma^\mu P_L d_j) 
  (\bar\ell \gamma_\mu \gamma_5 \ell')\,, \nonumber \\
&& \mathcal{O}_S^{\ell\ell'} \,= \,\frac{e^2}{(4\pi)^2}\, (\bar d_i P_R d_j)
  (\bar\ell  \ell')\,, \quad
\mathcal{O}_P^{\ell\ell'} \,= \,\frac{e^2}{(4\pi)^2}\, (\bar d_i P_R d_j)
  (\bar\ell \gamma_5 \ell')\,, \nonumber\\
&& \mathcal{O}^{\ell\ell'}_T \,= \,\frac{e^2}{(4\pi)^2}\, (\bar d_i
\sigma_{\mu\nu} d_j) 
  (\bar\ell \sigma^{\mu\nu}\ell')\,, \quad
\mathcal{O}^{\ell\ell'}_{T5} \,= \,\frac{e^2}{(4\pi)^2}\, (\bar d_i
\sigma_{\mu\nu} d_j) (\bar\ell \sigma^{\mu\nu} \gamma_5\ell')\,.
\end{eqnarray}
In the above, $e$ is the electric charge and 
$\sigma_{\mu\nu}=i[\gamma_\mu,\gamma_\nu]/2$. 
The set of primed operators ($X'=7',9',10',S',P'$) comprises 
those of opposite chirality, and can be obtained by replacing 
$P_L \leftrightarrow P_R$ in the quark currents.
The contribution of the right-handed current operators is negligible
in the SM. Flavour universality of lepton-gauge interactions in the SM 
implies that the Wilson coefficients of operators 
$\mathcal{O}_{i}^{\ell\ell}$ are universal for all lepton flavours 
$(\ell = e, \mu ,\tau)$, and the strict conservation of individual
lepton flavour further precludes  
cLFV Wilson coefficients $C_{i}^{\ell\ell'}$ $(\ell\ne \ell^\prime)$.

In the present scalar leptoquark model, once the heavy degrees of
freedom have been integrated out (under the assumption that 
$M_{t, W,Z}^2 \ll m_{h_1}^2$), the NP effective Hamiltonian is given 
by~\cite{Hiller:2017bzc}
\begin{equation}\label{eq:NPeffH_RK}
\mathcal{H}^\text{NP}(b\to s \ell^-\ell^{\prime +}) 
\,= \,- \frac{y_{b\ell^\prime} \,y_{s\ell}^\ast}{m_{h_1}^2}\,
(\bar s \,\gamma^\mu \,P_L \,b)\,(\bar\ell \,\gamma_\mu \,P_L\,\ell')+
\text{H.c.}\,; 
\end{equation}
comparing the above with the operator basis of
Eq.~(\ref{eq:effHbtos}), it is possible to infer the following
contributions to the Wilson coefficients
\begin{equation}\label{eq:bsll}
C_9^{\ell\ell^\prime} \,= \,-C_{10}^{\ell\ell^\prime}\, =\, 
\frac{\pi \,v^2}{\alpha_e \,V_{tb}\,V_{ts}^\ast} \,
\frac{y_{b\ell^\prime}\, y_{s\ell}^\ast}{m_{h_1}^2}\,.	
\end{equation}

The deviations from the SM lepton flavour universality imply that the
modifications to the Wilson coefficients are necessarily non-universal
for the muon and electron entries; the model-independent fit 
of~\cite{Hiller:2017bzc} suggests the following corrections 
at the $1\sigma$ range: 
\begin{align}\label{eq:RKstfit}
\text{Re}[C_{9, \text{NP}}^{\mu \mu}\,-\,C_{10, \text{NP}}^{\mu \mu} 
- (\mu \leftrightarrow e)] &\sim\, -1.1 \pm 0.3\,,\\ 
\text{Re}[C_{9^\prime}^{\mu \mu}-C_{10^\prime}^{\mu \mu} \,-\, 
(\mu \leftrightarrow e)] &\sim \,0.1\pm 0.4\,,
\end{align}
(notice that the second constraint is compatible with zero, and can
be fulfilled by setting $C_{9^\prime,10^\prime}^{\mu \mu, ee}=0$).
In the present NP construction, leptoquark couplings to both muons and
electrons are present, and are of left-handed nature. 
Given that $C_9^{\ell\ell^\prime}=-C_{10}^{\ell\ell^\prime}$
(cf. Eq.~(\ref{eq:bsll})), the best fit to the $C_{9,10 \text{ NP}}^{\mu
  \mu, ee}$ Wilson coefficients of Eq.~(\ref{eq:bsll}) can be recast
as 
\begin{equation}\label{eq:mu.e.fit}
-1.4 \, \lesssim  \,
2 \,\text{Re}[C_{9, {\rm NP}}^{\mu \mu}\,- \,C_{9, {\rm NP}}^{e e}]
 \,\lesssim \, -0.8\,.
\end{equation}
Global fits to a large number of observables probing 
lepton flavour universality in relation to 
$b\to s \mu^\pm \mu^\mp$, $b\to s e^\pm e^\mp$ and $b\to s
\gamma$ processes also suggest NP scenarios consistent with
the above fit to the LFUV $R_{K^{(*)}}$ observables.
A common conclusion that can be generically drawn is that the 
NP responsible for the observed discrepancies in $b\to s $ data
appears to predominantly couple to muons, and is strongly manifest in
vector operators (as ${\cal O}_{9,10}^{\mu\mu}$). 
Recent studies and fits by a number of authors
(see, e.g.~\cite{Descotes-Genon:2015uva,Hurth:2014vma,Altmannshofer:2014rta,Beaujean:2013soa,Capdevila:2017bsm}) 
advocate NP contribution to $C_9^{\mu\mu}$ ($\sim -1$) only,  
or then SU$(2)_L$ invariant scenarios
($C_9^{\mu\mu}=-C_{10}^{\mu\mu} \sim -0.6$)
as preferred NP solutions to alleviate the tensions with the SM. 

\begin{figure}[t!]
\begin{center}
\includegraphics[width=0.50\textwidth]{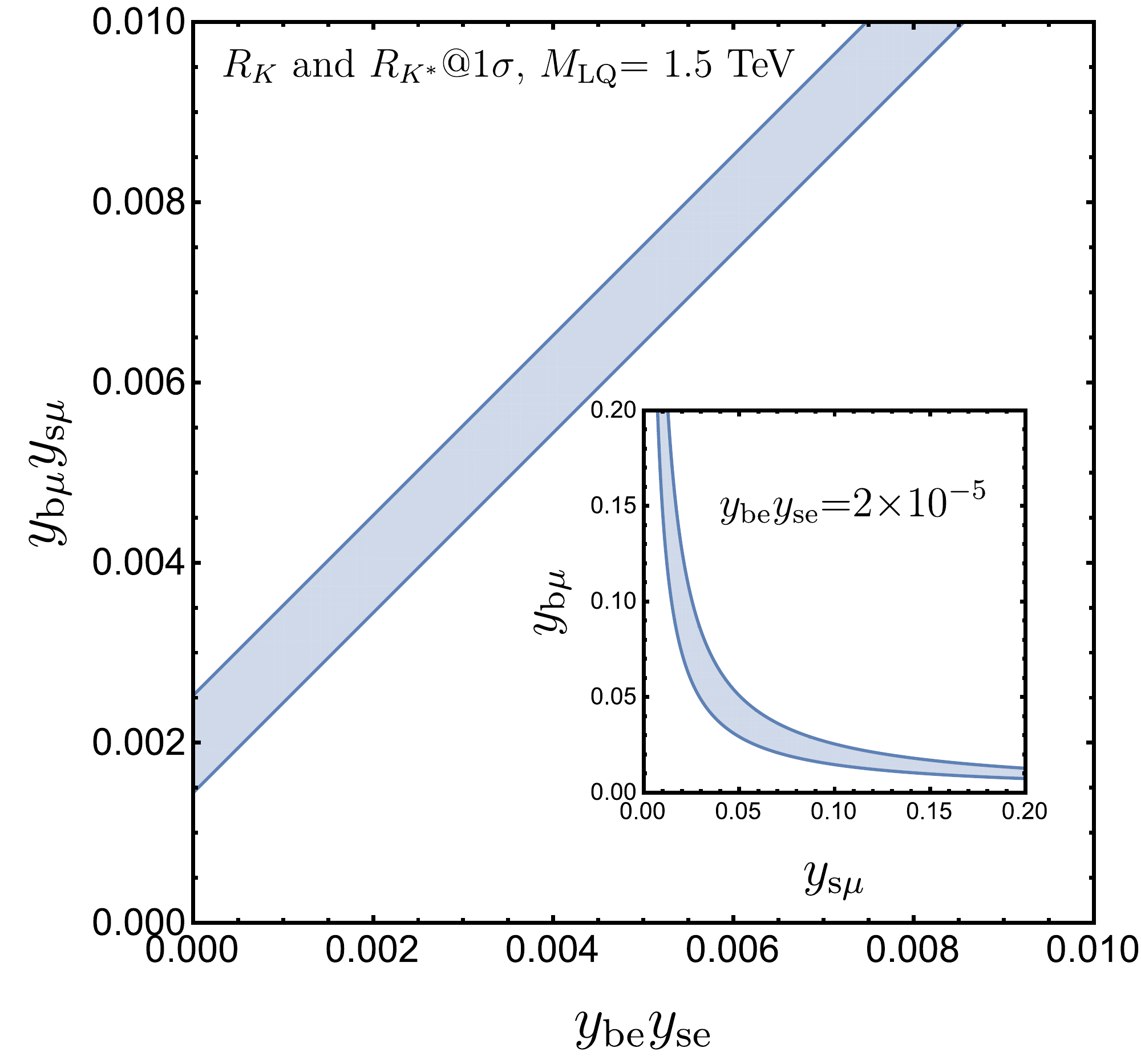}
\caption{On the main plot, 1$\sigma$ region in the 
  $(y_{be}\,y_{se}-y_{b\mu}\,y_{s\mu})$
  parameter space consistent with $R_K$ and $R_{K^\ast}$ data, 
  cf. Eq.~(\ref{eq:mu.e.coupling.fit}), for an example with 
  $m_{h_1} = 1.5~\text{TeV}$. 
  On the inset plot we display the
  $R_{K^{(\ast)}}$ allowed region in the $(y_{b\mu}-y_{s\mu})$
  plane, for fixed values of the leptoquark electron Yukawa
  couplings $y_{be} y_{se} = 2\times 10^{-5}$ (again for $m_{h_1} =
  1.5~\text{TeV}$). }  
\label{fig:RK}
\end{center}
\end{figure}

\bigskip
In terms of the $h_1$ leptoquark mass and couplings, 
the expression of Eq.~(\ref{eq:mu.e.fit}) translates into the
following condition~\cite{Hiller:2017bzc}
\begin{equation}\label{eq:mu.e.coupling.fit}
0.64 \times 10^{-3}\,\lesssim\, 
\frac{\text{Re}[y_{b\mu}\,y_{s\mu}^\ast \,-\, y_{be}\,y_{se}^\ast]}
{(m_{h_1}/1\text{TeV})^2}\, \lesssim \,1.12 \times 10^{-3}\,.
\end{equation}
In Fig.~\ref{fig:RK}, for an illustrative value of 
$m_{h_1} = 1.5~\text{TeV}$, 
we display the $(y_{b\ell}\,y_{s\ell})$ parameter space  
consistent with the observed values of $R_K$ and $R_{K^\ast}$ at the
$1\sigma$ level, in agreement with Eq.~(\ref{eq:mu.e.coupling.fit}). 
The inset plot shows the regimes of $y_{b\mu}$ and $y_{s\mu}$ 
compatible with $R_{K^{(\ast)}}$ (also for $m_{h_1} = 1.5~\text{TeV}$,
and for fixed values of the couplings to the electron, 
$y_{be} y_{se} = 2\times 10^{-5}$).

%

\subsection{Anomalies in $\pmb{b\to c \ell^-\bar \nu_i}$:
  $\pmb{R_{D^{(*)}}}$}\label{sec:RD*}

Several experimental collaborations have also reported deviations from
lepton flavour universality in association with $B \to D^{\ast} \ell
\bar \nu$ decays (charged current $b\to c \ell \bar \nu_i$
transitions).  
A scalar charged leptoquark with non-trivial couplings to quarks 
and leptons can mediate $d_k\to u_j \ell \bar \nu_i$ transitions 
at the tree level, via the exchange of a charged $h_1^{1/3}$. 
The SM effective Hamiltonian governing these transitions is thus
modified as follows:
\begin{align}\label{eq:effHRD}
\mathcal{H}_\text{eff}(d_k\to u_j\ell\bar \nu_i) \,&=\, 
\mathcal{H}_\text{eff}^\text{SM} \,+\,
\mathcal{H}_\text{eff}^\text{NP}
\nonumber \\
&=\, \left[\frac{4 \,G_F}{\sqrt{2}} \,V_{jk}\, U_{\ell i} \,-\, 
\frac{(y\,U)_{ki}\,(V\,y^\ast)_{j\ell}}{2 \,m_{h_1}^2}\right]\,
\left(\bar u_j\,\gamma^\mu \,P_L\, d_k\right)\,
\left(\bar \ell \,\gamma_\mu \,P_L\, \nu_i\right) + \,\text{H.c.}\nonumber\\
&= \,\frac{4 \,G_F}{\sqrt{2}}\,V_{jk}\,\left[ U_{\ell i} \,-\, 
\frac{v^2}{4 \,V_{cb}\,
  m_{h_1}^2}\,(yU)_{ki}\,(Vy^\ast)_{j\ell}\right]\,
\left(\bar u_j\,\gamma^\mu \,P_L\, d_k\right)\,
\left(\bar \ell \,\gamma_\mu \,P_L\, \nu_i\right)+ \text{H.c.}\;.
\end{align} 
The experimentally measured decay probability is an incoherent sum over the
(untagged) neutrino flavour  $i$; one thus finds
\begin{align}\label{eq:amp-sq-RD}
\left|\mathcal{A}(d_k\to u_j\ell\bar \nu)\right|^2 
&= \,\sum_i \left|\mathcal{A}_\text{SM}\right|^2 \,
\left| U_{\ell i}  \,-  \,\frac{v^2}{4  \,V_{jk}\,m_{h_1}^2}
 \,(yU)_{ki} \,(Vy^\ast)_{j\ell}\right|^2 \nonumber\\
=& \, \left|\mathcal{A}_\text{SM}\right|^2  \,
\left[1  \,+ \, |x_{j\ell}|^2 \,\sum_i |y_{ki}|^2  \,- \, 
2 \,\text{Re}\left(x_{j\ell}\,y_{k\ell}\right)\right]\,,
\end{align} 
where $x_{j\ell} = (Vy^\ast)_{j\ell}\left(v^2/4
V_{cb}m_{h_1}^2\right)$, and in which one has used the unitarity 
of the PMNS matrix.  
The SM width for the decay $B \to D^{\ast} \ell \bar \nu$
(i.e., for $d_k = b,\, u_j = c$)
will thus be corrected by an overall factor
\begin{equation}\label{eq:decaywidthRD}
\Gamma (B \,\to \,D^{\ast}\, \ell \,\bar \nu) \,= \,
\Gamma_\text{SM} (B \,\to\, D^{\ast}\, \ell \,\bar \nu) \,
\left[1 \,+\, |x_{c\ell}|^2\,\sum_{i=e,\mu,\tau} |y_{bi}|^2 \,-\, 
2 \,\text{Re}\left(x_{c\ell}\,y_{b\ell}\right)\right]\,.
\end{equation}
Pure leptoquark contributions are suppressed by an additional
$v^2/m_{h_1}^2$ factor with respect to the SM-NP interference term,
and can be hence neglected as a first approximation.

Defining $R_{D^{(\ast)}}$ as the ratio of the decay widths of tau and
muon modes - that is, 
$R_{D^{(\ast)}} = \text{BR}(B \to D^{(\ast)}\tau\bar \nu)/
\text{BR}(B \to D^{(\ast)}\mu\bar \nu)$, 
one can further construct the double ratio 
\begin{equation}\label{eq:RDoverRDSM-th}
\frac{R_{D}}{R_{D,\,\text{SM}}} \,= \,
\frac{R_{D^\ast}}{R_{D^\ast,\,\text{SM}}}\,=\, 
\frac{1 - 2 \,\text{Re}\left(x_{c\tau}\,y_{b\tau}\right)}
{1 - 2 \,\text{Re}\left(x_{c\mu}\,y_{b\mu}\right)}\,,
\end{equation}
(equal to one in the absence of NP).
After combining current experimental world averages with the SM
predictions, the current anomalous data can be parametrised 
as
\begin{equation}
\frac{R_{D}}{R_{D,\,\text{SM}}} \,= \,1.36 \pm 0.15\, , \quad
\frac{R_{D^{\ast}}}{R_{D^{\ast},\,\text{SM}}} \,= \,1.21 \pm 0.06\, , 
\end{equation}
in which the statistical and systematical errors have been added in
quadrature.
Similar ratios comparing distinct final state lepton flavours can
be built to test possible LFUV in the corresponding sectors. 
For example, the Belle Collaboration has reported measurements of the
ratios $R_{D^{(\ast)}}^{\mu/e} = \text{BR}(B \to D^{(\ast)}\mu\bar
\nu)/\text{BR}(B \to D^{(\ast)}e\bar \nu)$, which probe lepton
flavour universality between electron and muon modes. 
The experimental 
values $R_{D}^{\mu/e,\text{exp}} = 0.995\pm 0.022\pm 
0.039$~\cite{Glattauer:2015teq} and $R_{D^\ast}^{e /\mu,\text{exp}} =
1.04\pm 0.05\pm 0.01$~\cite{Abdesselam:2017kjf} are consistent with
the SM expectation, $\sim 1$. The averaged value of both measurements
is $R_{D^{(\ast)}}^{\mu/e,\text{exp}} = 0.977\pm 0.043$. 

In the present leptoquark model, the ratio $R_{D^{(\ast)}}^{\mu/e}$ is
given by the appropriately modified version of
Eq.~(\ref{eq:RDoverRDSM-th}), 
\begin{equation}\label{eq:RDmue}
\frac{R_{D^{(\ast)}}^{\mu/e}}{R_{D^{(\ast)}, \text{ SM}}^{\mu/e}} \,
=\, \frac{1\, -\, 2 \,\text{Re}\left(x_{c\mu}\,y_{b\mu}\right)}{1 
\,-\, 2 \,\text{Re}\left(x_{ce}\,y_{be}\right)}\,.
\end{equation}

After having detailed the leptoquark contributions to the 
meson observables 
currently exhibiting a significant deviation from the SM expectation, 
we now address other processes which - being in agreement with SM
predictions (negative searches or compatible measurements) can
constrain the masses of the new states and their couplings.

\section{Constraints from rare meson decays and
  oscillations}\label{sec:mesoncon} 

Various observables involving mesons lead to important 
bounds on leptoquark couplings; here we discuss the (leptoquark) 
NP contributions
to leptonic and semi-leptonic meson decays (occurring at the tree
level), and to meson oscillations and rare radiative decays, both at
the loop level. The SM predictions and current experimental bounds for
the processes here discussed are summarised in
Table~\ref{tab:flavour}.
The stringent bounds on NP contributions arising from the now observed
decay $B_s \to \mu^+ \mu^-$ are not discussed here, as they
have been implicitly taken into account in defining the allowed ranges
for $C_{10,\text{NP}}^{\mu\mu}$ and the $y_{22,23}$ Yukawa  
couplings in the previous section.
Likewise, we do not include
the constraints arising from semileptonic $K$-decays into charged dileptons 
($K \to \pi \ell \ell$), as theoretical (SM) predictions are plagued
by important uncertainties, and are not yet up to par with the precision 
of the experimental results (see for example \cite{Crivellin:2016vjc,Fajfer:2018bfj}).

{\small
\renewcommand{\arraystretch}{1.1}
\begin{table}[h!]
\begin{center}
\begin{tabular}{|c|c|c|}
\hline
Observables & SM prediction & Experimental data  \\
\hline
$\text{BR}(K^+ \to \pi^+ \nu\bar \nu)$ & 
$(8.4 \pm 1.0) \times 10^{-11} 
\phantom{|}^{\phantom{|}}_{\phantom{|}}$\cite{Buras:2015qea} & 
\begin{tabular}{l}
$17.3^{+11.5}_{-10.5} \times 10^{-11} 
\phantom{|}^{\phantom{|}}_{\phantom{|}}$\cite{Artamonov:2008qb}
\\
$< 11 \times 10^{-10} 
\phantom{|}^{\phantom{|}}_{\phantom{|}}$\cite{Na62:2018}
\end{tabular}
\\
\hline
$\text{BR}(K_L \to \pi^0 \nu\bar \nu)$ & 
$(3.4 \pm 0.6) \times 10^{-11} 
\phantom{|}^{\phantom{|}}_{\phantom{|}}$\cite{Buras:2015qea} & 
$ \leq  2.6 \times 10^{-8} 
\phantom{|}^{\phantom{|}}_{\phantom{|}}$\cite{Ahn:2009gb}\\
\hline
\begin{tabular}{l}
$R_K^{\nu\nu}$, $R_{K^\ast}^{\nu\nu}$
$\phantom{a}^{\phantom{|}}_{\phantom{|}}$
\\
($B\to  K^{(\ast)} \nu\bar \nu$)
\end{tabular} & 
$R_{K^{(\ast)}}^{\nu\nu}=1$ 
& 
\begin{tabular}{l}
$R_K^{\nu\nu} < 3.9$
$\phantom{a}^{\phantom{|}}_{\phantom{|}}$\cite{Grygier:2017tzo}
\\
$R_{K^\ast}^{\nu\nu} < 2.7 
\phantom{|}^{\phantom{|}}_{\phantom{|}}$\cite{Grygier:2017tzo}
\end{tabular}
\\
\hline
\begin{tabular}{c}
$B_s^0 - \bar B_s^0$ $\phantom{a}^{\phantom{|}}_{\phantom{|}}$
\\
{(mixing parameters)}
\end{tabular} & 
\begin{tabular}{l}
$\Delta_{s} = |\Delta_{s}|e^{i\phi_s}=1$ $\phantom{a}^{\phantom{|}}_{\phantom{|}}$
\\
$\phi_{s}=0$
\end{tabular}
&
\begin{tabular}{l}
$|\Delta_s|=1.01^{+0.17}_{-0.10} 
\phantom{|}^{\phantom{|}}_{\phantom{|}}$\cite{Charles:2015gya},\\
$\phi_s [^{\circ}]=1.3^{+2.3}_{-2.3} 
\phantom{|}^{\phantom{|}}_{\phantom{|}}$\cite{Charles:2015gya}
\end{tabular}
\\
\hline
\begin{tabular}{c}
$K^0 - \bar K^0$ $\phantom{a}^{\phantom{|}}_{\phantom{|}}$\\
$\Delta m_K/(10^{-15}\text{GeV})$
\end{tabular}
& 
\begin{tabular}{c}
\\
$3.1(1.2) 
\phantom{|}^{\phantom{|}}_{\phantom{|}}$\cite{Brod:2011ty}
\end{tabular}
&
\begin{tabular}{c}
\\
$3.484(6) 
\phantom{|}^{\phantom{|}}_{\phantom{|}}$\cite{Patrignani:2016xqp}
\end{tabular}
\\
\hline
$\text{BR}(K_L \to \mu e)$ & 
--- & 
$< 4.7\times 10^{-12} 
\phantom{|}^{\phantom{|}}_{\phantom{|}}$\cite{Patrignani:2016xqp}\\
\hline
$\text{BR}(B_s \to \mu e)$ & 
--- & 
$< 1.1 \times 10^{-8} 
\phantom{|}^{\phantom{|}}_{\phantom{|}}$\cite{Patrignani:2016xqp}\\
\hline
\end{tabular}
\caption{Relevant observables and current
  experimental bounds for leptonic and semi-leptonic meson decays and
  neutral meson anti-meson oscillations discussed in this section.
}\label{tab:flavour} 
\end{center}
\end{table}
\renewcommand{\arraystretch}{1.}
}

\subsection{Rare $\pmb{K}$ and $\pmb{B}$ meson decays: 
$\pmb{K \to \pi \nu_\ell\bar \nu_{\ell^\prime}}$ and $\pmb{B\to  K^{(\ast)}
  \nu_\ell\bar \nu_{\ell^\prime}}$}

The $s\to d \nu\nu$
and $b\to s \nu\nu$ transitions provide some of the most important
constraints on NP scenarios aiming at addressing 
the anomalies in $R_{K^{(\ast)}}$ and $R_{D^{(\ast)}}$ data.

Following the convention of~\cite{Buras:2014fpa}, at the quark level 
the $|\Delta S|=1$ rare decays $K^+\,(K_L)\to \pi^+\,(\pi^0)
\,\nu_\ell \bar\nu_{\ell^\prime }$ and  $B\to  K^{(\ast)} \nu_\ell\bar
\nu_{\ell^\prime}$ can be described
by the following short-distance effective Hamiltonian for 
$d_j\to d_i \nu_\ell\bar \nu_{\ell^\prime}$ 
transitions~\cite{Bobeth:2017ecx,Bordone:2017lsy} 
\begin{eqnarray}\label{eq:eff-H-Ktopi}
\mathcal{H}_\text{eff}({d_j \to d_i} \nu_\ell\bar \nu_{\ell^\prime}) &=& 
-\frac{4 \,G_F}{\sqrt{2}} \,V_{ti}^\ast \,V_{tj}\,
\frac{\alpha_e}{2\,\pi}\,
\left[C_{L,ij}^{\ell\ell^\prime} \,\left(\bar d_i\,\gamma_\mu \,P_L\,
  d_j\right)\, \left(\bar \nu_\ell\,\gamma^\mu\,
  \,P_L\,\nu_{\ell^\prime}\right) \right. \,+\nonumber\\ 
&&\phantom{-\frac{4 \,G_F}{\sqrt{2}} \,V_{ti}^\ast \,V_{tj}\,
\frac{\alpha_e}{2\,\pi}\,
[}+\left.
C_{R,ij}^{\ell\ell^\prime} \,\left(\bar d_i\,\gamma_\mu \,P_R\,
d_j\right)\, \left(\bar \nu_\ell \,\gamma^\mu
\,P_L\nu_{\ell^\prime}\right) \right] \,+\, \text{H.c.}\, ,
\end{eqnarray}
in which $i,j$ denote the down-type flavour content of the final and
initial state meson, respectively. In the SM,  
only the lepton flavour conserving left-handed operator is present;
the associated Wilson coefficient
$C_{L,ij}^{\ell\ell}$ corresponding to a $d_j\to d_i$ transition is
given by 
\begin{equation}\label{eq:WC_sd}
C_{L,ij}^{\ell\ell,\text{SM}} \,=\, 
-\frac{1}{\sin^2\theta_{w}}\,\left(X_t^\text{SM} 
\,+\, \frac{V_{ci}^\ast \,V_{cj}}{V_{ti}^\ast \,V_{tj}}\, X_c^\ell\right)\,. 
\end{equation}
In the above,  
$X_t^\text{SM}$ and $X_c^\ell$ ($\ell = e, \mu, \tau$) 
are the loop functions associated with the (lepton flavour conserving) 
contributions from the top and charm quarks, respectively. 
In the SM, computations of the top loop function 
(including Next-to-Leading Order QCD 
corrections~\cite{Buchalla:1993bv,Misiak:1999yg,Buchalla:1998ba} and 
the full two-loop electroweak corrections~\cite{Brod:2010hi})
have led to the result $X_t^\text{SM} =
1.481(9)$~\cite{Buras:2015qea}; 
the charm loop functions $X_c^\ell$ have been computed at 
NLO~\cite{Buchalla:1998ba, Buchalla:1993wq} and at
Next-to-NLO~\cite{Buras:2005gr,Buras:2006gb}, 
and their numerical value is given by $\frac{1}{3} \sum_{\ell}X_c^\ell
/|V_{us}|^4 =0.365 \pm 0.012$.

Once the heavy leptoquark degrees of freedom have been integrated out, the 
general effective Hamiltonian describing the NP contribution 
to $d_j \to d_i\nu_\ell \bar \nu_{\ell^\prime}$ processes 
induced by $h_1$ is given by 
\begin{equation}\label{eq:NPeffHbtosnunu}
\mathcal{H}_\text{eff}^\text{LQ}(d_j \to d_i \nu_\ell\bar
\nu_{\ell^\prime}) \,
=  \,- \frac{(y \,U)_{j\ell^\prime} \,(y \,U)_{i\ell}^\ast}{2m_{h_1}^2} \,
\left(\bar d_i \,\gamma_\mu  \,P_L  \,d_j\right) \,
\left(\bar \nu_\ell \,\gamma^\mu  \,P_L \,\nu _{\ell^\prime}\right)\,,
\end{equation}
and comparison with Eq.~(\ref{eq:eff-H-Ktopi}) leads to the following 
NP Wilson coefficient, 
\begin{equation}\label{eq:WC_s_to_d}
C_{L,ij}^{\ell\ell^\prime,\,\text{LQ}} \,= \,
\frac{\pi \,v^2}{\alpha_e \, V_{ti}^\ast V_{tj}}\,
\frac{(yU)_{i\ell}^\ast\,(yU)_{j\ell^\prime}}{2 \,m_{h_1}^2}\,,
\end{equation}
in which we notice the presence of the PMNS matrix $U$, and the
possibility that the tree level NP contribution, via the exchange
of $h_1^{1/3}$, may lead to different lepton
flavours in the final state, i.e. $\ell \neq \ell^\prime$.

\subsubsection*{$\pmb{K \to \pi \nu_\ell\bar \nu_{\ell^\prime}}$ decays}
The branching fraction for $K^+\to \pi^+ \nu_\ell\bar
\nu_{\ell^\prime}$ 
(corresponding to setting $i=d, j=s$ in the previous discussion) 
is given by~\cite{Bobeth:2017ecx,Bordone:2017lsy}
\begin{equation}\label{eq:BR_Kplus}
\text{BR}(K^+ \,\to \,\pi^+ \,\nu_\ell\,\bar \nu_{\ell^\prime}) \,= \,
\frac{\kappa_{+}\,(1+\Delta_{\rm em})}{3}
\sum_{\ell,\ell^\prime= e,\mu,\tau}
\left|\frac{V_{ts}^\ast \,V_{td}}{|V_{us}|^5}\,
X_t^{\ell\ell^\prime}\,+\, 
\frac{ V_{cs}^\ast \,V_{cd}}{|V_{us}|}\,
\delta_{\ell\ell^\prime}\,
\left(\frac{X_c^\ell}{|V_{us}|^4}\,+\,
\Delta_{P_{c,u}^\ell}\right)\right|^2\,, 
\end{equation}
in which $\kappa_{+}=5.173(25)\times 10^{-11}\,(|V_{us}|/0.225)^8$, 
$\Delta_\text{em} = -0.003$ is the electromagnetic correction from photon 
exchanges and $\Delta_{P_{c,u}} = 0.04(2)$ denotes 
the long-distance contribution from light quark loops, 
computed in~\cite{Isidori:2005xm}. 
The loop function associated with NP exchanges,
$X_t^{\ell\ell^\prime}$, is given by 
\begin{equation}{\label{eq:XT}}
X_t^{\ell\ell^\prime}\, = \, 
X_t^\text{SM} \,\delta_{\ell\ell^\prime} 
\,-\, \sin^2\theta_{w} \, C_{L,sd}^{\ell\ell^\prime,\,\text{LQ}}\,.
\end{equation}
Notice from both Eqs.~(\ref{eq:BR_Kplus}, \ref{eq:XT}) that the SM 
top and charm contributions are lepton flavour conserving.
Experimentally, the measurement of the charged kaon decay
mode~\cite{Artamonov:2008qb}, 
\begin{equation}\label{eq:exp_BR_Kaon+}
\text{BR}(K^+ \,\to \,\pi^+ \,\nu\,\bar \nu)_\text{exp} \,= \,
17.3^{+11.5}_{-10.5} \times 10^{-11}\,, 
\end{equation}
is expected to be improved in the near future
by the results of the NA62 collaboration. 
The NA62 recent measurement~\cite{Na62:2018} (one event) 
$\text{BR}(K^+ \to \pi^+ \nu\bar \nu)_\text{exp}= 28^{+44}_{-23}
\times 10^{-11}$
still suffers from large statistical uncertainties, but these are
expected to improve by an order of magnitude in coming months.
The experimental bounds will allow to better constrain
$X_t^{\ell\ell^\prime}$, and hence the leptoquark contributions,
encoded in $C_{L,ij}^{\ell\ell^\prime,\,\text{LQ}}$.

\subsubsection*{$\pmb{K_L\to \pi^0 \nu_\ell \nu_{\ell^\prime}}$ decays}
The branching fraction for the neutral mode 
$K_L\to \pi^0 \nu_\ell \nu_{\ell^\prime}$ can be written 
as~\cite{Bobeth:2017ecx,Bordone:2017lsy}
\begin{equation}\label{eq:BR_Klong}
\text{BR}(K_L \,\to \,\pi^0 \,\nu_\ell\,\bar \nu_{\ell^\prime}) \,=\, 
\frac{\kappa_{L}}{3}\sum_{\ell,\ell^\prime= e,\mu,\tau}
\left[\text{Im}\left(\frac{V_{ts}^\ast V_{td}}{|V_{us}|^5}\,
X_t^{\ell\ell^\prime}\right)\right]^2\,,  
\end{equation}
with $\kappa_L = 2.231(13)\times 10^{-10}\,(|V_{us}|/0.225)^8$ 
in the framework of the SM.
Concerning the experimental status, only 
a $90\%$ C.L. bound has been reported for the
decay~\cite{Ahn:2009gb}
\begin{equation}\label{eq:exp_BR_Kaon0}
\text{BR}(K_L \,\to \,\pi^0 \,\nu\,\bar \nu)_\text{exp} \,\leq\, 
2.6 \times 10^{-8}\, ,
\end{equation}
which - as occuring for the charged decay modes - can also be used to 
constrain $X_t^{\ell\ell^\prime}$ and
$C_{L,ij}^{\ell\ell^\prime,\,\text{LQ}}$ (albeit leading to weaker
constraints than those inferred from the  
$K^+ \,\to \,\pi^+ \,\nu\,\bar \nu$ mode).

\subsubsection*{$\pmb{B\to  K^{(\ast)} \nu_\ell\bar \nu_{\ell^\prime}}$}
The branching ratio for the rare $B$ decay can be obtained by
considering the general expressions for the effective Hamiltonian and
NP Wilson coefficients, respectively given in Eqs.~(\ref{eq:eff-H-Ktopi},
\ref{eq:WC_s_to_d}), with $i=s, j=b$.
It proves convenient to consider the following ratios 
$R^{\nu\nu}_{K^{(\ast)}} = \Gamma_\text{SM + NP}(B\to
K^{(\ast)}\nu\nu)/\Gamma_\text{SM}(B\to K^{(\ast)}\nu\nu)$, 
which are then given by 
\begin{equation}\label{eq:Rnunu.0}
R^{\nu\nu}_{K} \,= \,R^{\nu\nu}_{K^{\ast}} \,=\,  
\frac{1}{3 \,|C_{L,bs}|^2}\sum_{\ell,\ell^\prime} 
\left|C_{L,bs}^{\ell\ell,\text{SM}}\,\delta_{\ell\ell^\prime}\,+\, 
C_{L,bs}^{\ell\ell^\prime,\text{LQ}}\right|^2\,,
\end{equation}
in which $C_{L, \text{SM}}^{\ell\ell} = -6.38(6)$ (for each neutrino
flavour)~\cite{Brod:2010hi}.
The above expression can be explicitly cast 
in terms of the leptoquark masses and couplings as follows:
\begin{eqnarray}\label{eq:Rnunu}
R^{\nu\nu}_{K^{(\ast)}} &=&  
\frac{1}{3 \,\left|C_{L,bs}^{\ell\ell,\text{SM}}\right|^2}\,
\left[3 \,\left|C_{L,bs}^{\ell\ell,\text{SM}}\right|^2 + 
\sum_{\ell\neq\ell^\prime}\left\lvert
\left(\frac{\pi \,v^2}{\alpha_e \,V_{tb}\, V_{ts}^\ast}\right) \,
\frac{y_{b\ell^\prime}\,y_{s\ell}^\ast}{2 \,m_{h_1}^2}
\right\rvert^2 \right. \,+\nonumber\\
&&\left.  
+ \sum_{\ell=\ell^\prime} \left\{\left\lvert
\left(\frac{\pi \,v^2}{\alpha_e \,V_{tb} \,V_{ts}^\ast}\right) \,
\frac{y_{b\ell^\prime}\,y_{s\ell}^\ast}{2 \,m_{h_1}^2}
\right\rvert^2+  
2\,\text{Re}\left[{C_{L,bs}^{\ell\ell,\text{SM}}}^\ast 
\left(\frac{\pi \,v^2}{\alpha_e \,V_{tb} \,V_{ts}^\ast}\right)\,
\frac{y_{b\ell}\,y_{s\ell}^\ast}{2 \,m_{h_1}^2}\right] 
\right\} \right]\,.
\end{eqnarray}
The SM predictions for each of the modes are 
BR$(B^+\to K^+ \nu\bar\nu) = (4.0\pm 0.5) 
\times 10^{-6}$~\cite{Buras:2014fpa} and
BR$(B^0\to K^{\ast 0} \nu\bar \nu) = (9.2\pm 1.0) 
\times 10^{-6}$~\cite{Buras:2014fpa}.

The latest experimental data from Belle~\cite{Grygier:2017tzo}  
allows to infer the following bound, 
$R^{\nu\nu}_{K^{(\ast)}} <3.9 (2.7)$ at $90\%$ C.L., which can be 
used to constrain combinations of leptoquark couplings
(lepton flavour conserving and violating) as given 
in Eq.~(\ref{eq:Rnunu}).

\subsection{Neutral meson mixings and oscillations}
Contributions to 
neutral meson mixings, $P-\bar P$  with $P= B^0_s, B^0_d ,K^0$,
arise both from SM box diagrams involving top and $W$'s, and from
NP box diagrams involving charged (neutral) leptons and 
$h_1^{4/3}(h_1^{1/3})$ leptoquarks. 
These contributions can be described
in terms of the following effective Hamiltonian for $|\Delta F| =2$ 
transitions
\begin{equation}\label{eq:PPmixing}
 \mathcal{H}_{P} \,= \,
(C_P^\text{SM}+C_P^\text{NP})
\left(\bar d_i\,\gamma^\mu \,P_L \,d_j\right)
\left(\bar d_i \,\gamma_\mu \,P_L\, d_j\right)\,+\text{H.c.},
\end{equation}
where~\cite{Dorsner:2017ufx}
\begin{equation}\label{eq:CNP}
C_P^\text{NP}=\frac{3}{128 \,\pi^2 \,m_{h_1}^2} 
\left({\sum_\ell} y_{i\ell}^\ast\,y_{j\ell}\right)^2 \,,
\end{equation}
with $\{i,j\}$ respectively denoting $\{b,s\}$, $\{b,d\}$ or $\{d,s\}$
for $P=B^0_s$, $B^0_d$ or $K^0$ mesons.

Let us begin by discussing $B^0_s$ mixing, which is potentially the 
process most sensitive to the couplings $y_{b\ell}$ and
$y_{s\ell}$ which are at the origin of the $R_{K^{(*)}}$ anomalies. 
Following~\cite{Lenz:2010gu,Charles:2015gya,Bona:2007vi}, 
one can define the ratio of the total contribution (NP and
SM) to the SM one
\begin{equation}  
\Delta_{s}\,=\,|\Delta_s|\, e^{ i \phi_{s} } \,=\, \frac{\langle B_s |
\mathcal{H}_{B_s} | \bar{B}_s \rangle}{\langle B_s |
\mathcal{H}^\text{SM}_{B_s} | \bar{B}_s \rangle}\,
=\,1\,+\,\frac{C_{B_s}^\text{NP}}{C_{B_s}^\text{SM}}
\,\equiv \,1\,+\,p_s\,. 
\end{equation}
In the leptoquark model we consider, the 
relative NP contribution $p_s$ can be cast as
\begin{equation}
p_{s} \,=\, 
\frac{ 3\,\left({\sum_\ell} \,y_{b\ell}\,y_{s\ell}^\ast\right)^2}{32\, 
m_{h_1}^2 \,G_F^2\,M_W^2\, S_0(x_t)\,(V_{tb} V^{\ast}_{ts})^2 } \,,
\end{equation}
where $S_0(x_t)=2.35$ is the Inami-Lim
function for the SM top quark box, 
with $x_t = M_t^2/M_W^2$~\cite{Inami:1980fz,Buras:1998raa}.
Current global fits are compatible with the 
SM value ($\Delta^\text{SM}_s=1$):
CKMfitter reports $|\Delta_s|=1.01^{+0.17}_{-0.10}$,
$\phi_s [^{\circ}]=1.3^{+2.3}_{-2.3}$~\cite{Charles:2015gya}, 
while for UTfit one has $|\Delta_s|=1.070\,{\pm 0.088}$,
$\phi_s[^{\circ}]=0.054\,{\pm0.951}$~\cite{Bona:2007vi}. 
Both analyses obtain their tightest $1\sigma$ constraints for
imaginary NP contributions (i.e., $\arg p_s=-\pi/2$): 
$|p_s|<0.016$. In our study, and for simplicity, we
will assume real Yukawa couplings $y_{q\ell}$ and a real $p_s$. In this
case one has $p_s\geq0$, and both analyses lead to
$p_s<p_s^{\text{max}}\approx 0.17$, which translates into 
the following $1\sigma$  upper
bound for the (real) $h_1$ leptoquark couplings:
\begin{equation}\label{eq:PPmbound}
\left|\sum_\ell y_{b\ell}\,y_{s\ell}\right| \,
\lesssim 0.079
\,\frac{m_{h_1}}{\text{TeV}}\,\sqrt{\frac{p_s^{\text{max}}}{0.17}}\,. 
\end{equation}
Following the same global analyses, a similar bound can be inferred 
from data on $B^0_d$ mixing (still in the case of real couplings),
which leads to $p_d<p_d^{\textrm{max}}\approx 0.13$. One thus has
\begin{equation}\label{eq:PPmBdbound}
\left|\sum_\ell y_{b\ell}\,y_{d\ell}\right| \,
\lesssim 0.069 
\,\frac{m_{h_1}}{\text{TeV}}\,\sqrt{\frac{p_d^{\text{max}}}{0.13}}\,.
\end{equation}

\bigskip
\noindent 
For the case of $K^0-\bar K^0$ mixing, 
the SM effective coupling can be expressed as
\begin{equation}\label{eq:CKSM}
C_K^\text{SM} \,= \,
\frac{G_F^2\,M_W^2}{4\pi^2}\,F^{*}(x_c, x_t)\,
\end{equation}
where $F(x_c, x_t)$ denotes the contribution of the 
distinct Inami-Lim functions, and is defined as  
\begin{equation}\label{eq:K0:IL}
F(x_c, x_t) \,= \,
(V_{cs}^{*} V_{cd})^2
\,\eta_{cc} \,S_0(x_c) \,+\, 
(V_{ts}^{*} V_{td})^2
\,\eta_{tt} \,S_0(x_t) \,+\,  
2\,V_{ts}^{*} V_{td}\,V_{cs}^{*} V_{cd}
\,\eta_{ct} \,S_0(x_c, x_t)\,,
\end{equation}
with $S_0(x_c)\approx x_c+O(x_c^2)=
m_c^2/M_W^2$~\cite{Inami:1980fz,Buras:1998raa}; the coefficients
$\eta_{cc} = 1.87(76)$, $\eta_{tt} =0.5765(65)$ and $\eta_{ct}
=0.496(47)$ encode
NNLO QCD corrections~\cite{Buras:1990fn, Brod:2011ty,
  Brod:2010mj,Buras:2015jaq}. 
The last two terms in Eq.~(\ref{eq:K0:IL}) can be safely neglected (as
they are CKM-suppressed); the first (and dominant) term is associated
to important theoretical uncertainties, $\mathcal{O}(40\%)$.
From Eq.~(\ref{eq:CNP}), the leptoquark contribution can be written as
\begin{equation}
p_{K} \,=\,\frac{C_K^\text{NP}}{C_K^\text{SM}}\,=\, 
\frac{ 3\,\eta_{tt}\,\tilde r\,\left({\sum_\ell} \,y_{s\ell}\,
y_{d\ell}^\ast\right)^2}{32\, m_{h_1}^2\, G_F^2\,M_W^2\, 
S_0(x_c)\eta_{cc}\,(V_{cs} V^{\ast}_{cd})^2 } \,,
\end{equation}
in which $\tilde r\approx 0.95$ allows to take 
into account the difference between the relevant scales 
($M_t$ and $m_{h_1}$)~\cite{Buras:2012dp}.

Real couplings $y_{q\ell}$ only affect $\Delta m_K =\text{Re}\langle
\bar K^0 |\mathcal{H}_{K} | K^0 \rangle/m_K$. Taking into account only
the (dominant) first term of Eq.~(\ref{eq:K0:IL}), 
the SM short distance contribution (cf. Eq.~(\ref{eq:CKSM})) 
gives~\cite{Brod:2011ty,Buras:2014maa}
\begin{equation}\label{eq:DeltamKcc}
(\Delta m_K)_\text{SM}=(3.1\pm 1.2)\times 10^{-15} \text{GeV} =
(0.89\pm0.34)\,(\Delta m_K)_\text{exp} 
\end{equation}
Comparing $(\Delta m_K)_\text{exp}$ with the theoretical prediction 
($(\Delta m_K)_\text{SM} +(\Delta m_K)_\text{NP}+(\Delta
m_K)_\text{LD}$), thus allows to obtain a 
conservative upper bound, 
$p_K<p_K^{\text{max}}\approx 0.45/0.55=0.81$.
Leading to the latter limit, one has 
taken the lowest values of both $(\Delta m_K)_\text{SM}$ ($\sim 55\%$) and  
the long distance (LD) 
contributions $(\Delta m_K)_\text{LD}$, which are hard to evaluate
but are expected to be positive like $p_k$~\cite{Buras:2014maa}. This
translates into an upper bound on the leptoquark couplings:
\begin{equation}\label{eq:PPmKbound}
\left|\sum_\ell y_{s\ell}\,y_{d\ell}\right| \,
\lesssim 0.014
\,\frac{m_{h_1}}{\text{TeV}}\,\sqrt{\frac{p_K^{\text{max}}}{0.81}}\,. 
\end{equation}

\subsection{Leptonic decays of pseudoscalar mesons 
$\pmb{P \to \ell^- \ell^{\prime+}}$}

The leptonic decays of pseudoscalar mesons are known to provide
stringent constraints on models of NP with modified lepton (and/or
quark) sectors; leptoquark models are no exception, and in what
follows we discuss some relevant modes of $K$ and $B$ mesons.  

Leptonic $B_s$ meson decays $B_{(s)}^0 \to \ell^\pm \ell^\mp$ 
are well predicted in the SM (the only hadronic uncertainty coming
from the decay constant $f_{B_{s}}$). 
At present, only the $B_s \to \mu^+ \mu^-$ decay mode has been 
observed, and it is in agreement with the SM. The LHC collaborations
have reported values for its branching fraction of 
$(2.8^{+0.7}_{-0.6})\times
10^{-6}$~\cite{CMS:2014xfa,Aaboud:2016ire}. As mentioned before, 
these bounds have been taken into account upon saturation of the $B$
decay anomalies.

For both $K$ and $B$ mesons, the cLFV leptonic decays have been shown
to lead to important constraints on NP models:
the cLFV $B$ decay modes are particularly relevant for leptoquark 
SM extensions~\cite{Becirevic:2016zri};
although the hard to quantify long-distance QCD corrections render
non-trivial an estimation of the leptonic $K_L$ 
decays~\cite{DAmbrosio:1997eof}), the 
upper bounds on the cLFV mode $K_L \to \mu^\pm e^\mp$ prove to be 
one of the most stringent constraints on the
couplings of leptoquarks to the first two generations of 
leptons and down-type quarks.

Following the effective Hamiltonian conventions adopted in
Eq.~(\ref{eq:effHbtos}), the decay width of  
$P \to \ell^\pm \ell^{\mp \prime}$ is governed by 
the $\mathcal{O}^{ij;\ell \ell^\prime}_{9}$ and 
$\mathcal{O}^{ij;\ell \ell^\prime}_{10}$ operators. In terms of the
corresponding Wilson coefficients
$C^{ij;\ell \ell^\prime}_{9,10}$, the decay width can be 
written~\cite{Becirevic:2016zri}
\begin{align}\label{eq:Pdecay}
\Gamma_{P \to \ell^{-}\ell^{\prime+}} &=\, 
f_{P}^2 \,m_{P}^3 \,\frac{G_F^2 \,\alpha_{e}^2}{64\,\pi^3}  
\left|V_{qj}\,V_{qi}^*\right|^2\,
\beta(m_{P},m_\ell,m_{\ell^{\prime}})\,\times \nonumber\\
& \times \Bigg[\left(
1-\frac{(m_\ell+m_{\ell^{\prime}})^2}{m_{P}^2}\right)
\left|\frac{m_{P}}{(m_{i} + m_{j})}(C^{ij;\ell \ell^{\prime}}_S 
-C^{ij;\ell \ell^{\prime}}_{S^\prime}) 
+ \frac{(m_\ell-m_{\ell^{\prime}})}{m_{P}} 
\left(C^{ij;\ell \ell^{\prime}}_{9} - C^{ij;\ell \ell^{\prime}}_{9^\prime} \right)
\right|^2 \,+ \nonumber\\
 & + \left(1-\frac{(m_\ell-m_{\ell^{\prime}})^2}{m_{P}^2}\right)
\left|\frac{m_{P}}{(m_{i} + m_{j})}(C^{ij;\ell \ell^{\prime}}_P -
C^{ij;\ell \ell^{\prime}}_{P^\prime}) + 
\frac{(m_\ell+m_{\ell^{\prime}})}{m_{P}} 
\left(C^{ij;\ell \ell^{\prime}}_{10} - C^{ij;\ell \ell^{\prime}}_{10^\prime} \right)
\right|^2 \Bigg]\,,
\end{align}
in which $m_{i,j}$ denotes the mass of the meson valence
quarks, the index $q$ refers
to up-type quarks (the sum being dominated by the top quark
contribution), and one defines
\begin{equation}
\beta(m_{P},m_\ell,m_{\ell^{\prime}}) \,=\, 
\sqrt{[1-(m_\ell-m_{\ell^{\prime}})^2/m_{P}^2]\, 
[1-(m_\ell+m_{\ell^{\prime}})^2/m_{P}^2)]}\,.
\end{equation}

\section{Charged lepton flavour violating processes }\label{sec:lfv}

Due to the presence of new states with non-negligible couplings to 
neutral and charged leptons, which are a source of LFUV, 
one expects that the model under consideration will
give rise to important contributions to cLFV observables.

While most cLFV decays correspond to higher order (loop) processes, it is
important to notice that transitions occurring in the presence of
matter, such as neutrinoless muon-electron conversion in nuclei
can now occur at the tree level. In the following, we address the
contributions of the model to several cLFV 
observables\footnote{In what concerns three-body decays we focus our
  discussion on the case of same-flavour final lepton state, i.e., 
$\ell \to 3 \ell^\prime$.}, whose
experimental status (current bounds and future sensitivities) is
summarised in Table~\ref{tab:cLFV}. The bounds on 
leptonic observables will give rise to stringent constraints on the
parameter space of the model: 
other than neutrino oscillation data, bounds on
several lepton flavour violating observables will play a significant
r\^ole in identifying the regimes which can successfully lead to an
explanation of the $B$ meson anomalies. 

{\small
\renewcommand{\arraystretch}{1.1}
\begin{table}[h!]
\begin{center}
\begin{tabular}{|c|c|c|}
\hline
cLFV process & Current experimental bound & Future sensitivity   \\	
\hline
\hline
$\text{BR}(\mu\to e \gamma)$	&  
$ 4.2\times 10^{-13}$ (MEG~\cite{TheMEG:2016wtm})	&  
$6\times 10^{-14}$ (MEG II~\cite{Baldini:2018nnn}) \tablefootnote{For a relevant discussion concerning the future perspectives of the searches for $\mu\rightarrow e \gamma$ process see \cite{Cavoto:2017kub}.}  	\\
$\text{BR}(\tau \to e \gamma)$	& 
$ 3.3\times 10^{-8}$ (BaBar~\cite{Aubert:2009ag})	 & 
$10^{-9}$ (Super B~\cite{Aushev:2010bq}) 	 	\\
$\text{BR}(\tau \to \mu \gamma)$	& 
$ 4.4\times 10^{-8}$ (BaBar~\cite{Aubert:2009ag})	 & 
$10^{-9}$ (Super B~\cite{Aushev:2010bq})		\\
\hline
\hline
$\text{BR}(\mu \to 3 e)$	&  
$1.0\times 10^{-12}$ (SINDRUM~\cite{Bellgardt:1987du}) 	& 
$10^{-15(16)}$ (Mu3e~\cite{Blondel:2013ia})  	\\
$\text{BR}(\tau \to 3 e)$	&  
$2.7\times 10^{-8}$ 	(Belle~\cite{Hayasaka:2010np})& 
$10^{-9}$ (Super B~\cite{Aushev:2010bq})  	\\
$\text{BR}(\tau \to 3 \mu )$	& 
$3.3\times 10^{-8}$ (Belle~\cite{Hayasaka:2010np})	 & 
$10^{-9}$ (Super B~\cite{Aushev:2010bq})		\\
\hline
\hline
$\text{CR}(\mu- e, \text{N})$ & 
$ 7 \times 10^{-13}$ (Au, SINDRUM~\cite{Bertl:2006up}) & 
$10^{-14}$ (SiC, DeeMe~\cite{Nguyen:2015vkk})    \\
& & $10^{-15 (-17)}$ (Al, COMET~\cite{Krikler:2015msn})  \\
& & $3 \times 10^{-17}$ (Al, Mu2e~\cite{Bartoszek:2014mya}) \\
& & $10^{-18}$ (Ti, PRISM/PRIME~\cite{Kuno:2005mm})  \\
\hline
\hline
\end{tabular}
\caption{Current experimental bounds and 
future sensitivities of cLFV processes included in the analysis.}
\label{tab:cLFV}
\end{center}
\end{table}
\renewcommand{\arraystretch}{1.}
}

\subsection{Radiative decays $\ell \to \ell^\prime \gamma$}

In the present model, radiative charged lepton decays are induced at
the loop level ($h_1$ leptoquarks and quarks running in the loop). 
The effective Lagrangian for $\ell \to \ell^\prime \gamma$ decays 
can be written as \cite{Dorsner:2016wpm}
\begin{equation}\label{eq:l-lprimegamma}
\mathcal{L}_\mathrm{eff}^{\ell \to \ell^\prime \gamma} 
\,= \,-\frac{4\,G_F}{\sqrt{2}}\,
\bar\ell^\prime \sigma^{\mu\nu} F_{\mu\nu}  
\left(C_{L}^{\ell\ell'} P_L + C_{R}^{\ell\ell'} P_R \right) \ell
\, +\, \text{H.c.}\,,
\end{equation}
with $F_{\mu \nu}$ the electromagnetic field strength.
The flavour violating coefficients can be cast as
\begin{equation}{\label{eq:Cnorm}}
C_{L (R)}^{\ell\ell'} \,= \, 
\frac{e}{4\sqrt{2}G_F} \sigma^{\ell\ell'}_{L(R)}\,;
\end{equation}
the new states and interactions give rise to  
the following effective coefficients 
$\sigma^{\ell\ell'}_{L(R)}$, 
\begin{align}
\label{eq:llgamma-loops}  
& \sigma^{\ell\ell'}_{L}\, = \,\frac{3}{16\,\pi^2 \,m_{h_1}^2} 
\sum_q X_{q\ell'}^* \, X_{q\ell} \,m_{\ell'}\,  
\left[Q_S \,f_S(x_q) \,- \,f_F(x_q)\right]\,, \nonumber\\
& \sigma^{\ell\ell'}_{R} \,= \,\frac{3}{16\,\pi^2 \,m_{h_1}^2} 
\sum_q  X_{q\ell'}^* \,X_{q\ell} \,m_{\ell} \,
\left[Q_S \,f_S(x_q) \,-\, f_F(x_q)\right]\, .
\end{align}
In the above, 
$X_{q\ell}\equiv -\sqrt{2} y_{q\ell}$ and $Q_S=4/3$ for down-type
quarks ($q=d$), 
while $X_{q\ell}\equiv -V_{qq'}^{*}y_{q'\ell}$ and $Q_S=1/3$ 
for up-type quarks ($q=u$). The loop functions, cast in terms of 
$x_q = m_q^2/m_{h1}^2$, are given by
\begin{equation}
\label{eq:llgamma-func}
  f_S(x) \,=\, \frac{x+1}{4\,(1-x)^2} + \frac{x \,\log x}{2\,(1-x)^3} ,\quad
  f_F(x) \,= \,\frac{x^2-5x-2}{12\,(x-1)^3} + \frac{x\, \log x}{2\,(x-1)^4}\,.
\end{equation}
Finally, the cLFV radiative decay width is given by
\begin{equation}
\Gamma(\ell \to \ell^\prime \gamma) \,= \,\frac{\alpha_{e}\, m_\ell^3 
\,\left(1-m_{\ell'}^2/m_\ell^2\right)^3}{4}
\left(|\sigma^{\ell\ell'}_L|^2 \,+\, |\sigma^{\ell\ell'}_R|^2 \right). 
\end{equation}

\subsection{Three body decays $\ell \to \ell^\prime \ell^\prime
  \ell^\prime$}

The photonic interactions at the source of the radiative decays
(parametrised by the couplings $C_{L,R}^{\ell\ell'}$, see 
Eq.~(\ref{eq:l-lprimegamma})) will also induce the three-body cLFV
decays; moreover, direct four-fermion interactions
are responsible for additional contributions. 

Following~\cite{Okada:1999zk,Kuno:1999jp}, the low-energy effective
Lagrangian including the four-fermion (contact) operators 
responsible for $\ell \to \ell'\ell' \ell'$ decays can be written as  
\begin{eqnarray}\label{eq:lto3l}
\mathcal{L}_{\ell \to \ell' \ell' \ell'} &=&
-\frac{4\,G_F}{\sqrt{2}} \left[ 
g_1 \,(\bar {\ell'}\,P_L\, \ell) (\bar {\ell'}
\,P_L\, \ell')\,+\,g_2 \,(\bar {\ell'} \,P_R\, \ell) (\bar {\ell'}\,
P_R \,\ell') \right. \,+\nonumber\\
&& \left.\,+\,g_3 \,(\bar {\ell'} \,\gamma^\mu \,P_R \,\ell) (\bar {\ell'}
\, \gamma_\mu \,P_R \,\ell')\,+\,
g_4\, (\bar {\ell'} \,\gamma^\mu \,P_L \,\ell) (\bar {\ell'}
\,\gamma_\mu \,P_L\, \ell') \,+\,\right. \nonumber\\
&& \left. \,+\, g_5\, (\bar {\ell'} \,\gamma^\mu \,P_R\, \ell) (\bar {\ell'}
\, \gamma_\mu \,P_L \,\ell') 
\,+\, g_6 \,(\bar {\ell'} \,\gamma^\mu \,P_L \,\ell) (\bar {\ell'}
\,  \gamma_\mu \,P_R\, \ell') \right]\, +\, \text{H.c.}\,.
\end{eqnarray}
In the model under study, there are several types of diagrams
contributing to the 3-body cLFV decays:  photon penguins (dipole and
off-shell ``anapole''), $Z$ penguins and box diagrams, all due to
flavour violating interactions involving the scalar leptoquark
$h_1$ and quarks. 
Neglecting Higgs-mediated exchanges,
the distinct diagrams will give rise to non-vanishing contributions to
the dipole operators ($C_{L,R}^{\ell\ell'}$), as well as to $g_3$, 
$g_4$, $g_5$ and $g_6$.
The box and the $Z$ penguin diagrams contribute to $g_4$ and $g_6$ 
as follows~\cite{Dorsner:2017ufx}
\begin{align}
g_4^{\text{box},Z} & =\, 
\frac{\sqrt{2}}{4\,G_F} \,
\frac{3 \,(y^\dagger \,y)_{\ell'\ell}}{(4\,\pi)^2 \,m_{h_1}^2}\,
\left[ (y^\dagger \,y)_{\ell'\ell'} \,+\,
\frac{\sqrt{2}}{9} \,G_F \,M_W^2 \,(2-3 \cos^2\theta_w -
3\log x-3\pi i)\right]\,,\nonumber\\
g_6^{\text{box},Z} & = \,\frac{\sqrt{2}}{4\,G_F} 
\frac{3 \,(y^\dagger \,y)_{\ell'\ell}}{(4\,\pi)^2 \,m_{h_1}^2}
\frac{2\,\sqrt{2}}{9} \,G_F\, M_Z^2 \,\sin^2 \theta_w \,(2-3
\cos^2\theta_w -3\log x-3\pi i)\,,
\end{align}
in which $x = M_Z^2/m_{h_1}^2$. 
The off-shell $\gamma$-penguin diagrams induce non-vanishing
contributions to $g_{3,5}$ and $g_{4,6}$, which are given 
by~\cite{Kuno:1999jp}
\begin{align}
& g_3^{\gamma} \,= \,g_5^{\gamma} \,=\,  
\frac{\sqrt{2} \,e^2}{4\,G_F \,m_\mu^2}\,
\left[\tilde{f}_{E0}(0) \,+\, \tilde{f}_{M0}(0)\right]\, ,
\nonumber \\
& g_4^{\gamma} \,=\, g_6^{\gamma}\,=\, 
\frac{\sqrt{2} \,e^2}{4\,G_F\,m_\mu^2}\,
\left[\tilde{f}_{E0}(0) \,-\, \tilde{f}_{M0}(0)\right]\,,
\end{align}
in which the form factors can be cast as
\begin{equation}\label{eq:mu_e_form} 
f_{E0}(q^2)\,=\, \frac{q^2}{m_\mu^2}\,\tilde{f}_{E0}(q^2)\,,
\quad \quad
f_{M0}(q^2)\,=\, \frac{q^2}{m_\mu^2}\,\tilde{f}_{M0}(q^2)\,,
\end{equation}
and are defined in such a way that $\tilde{f}_{E0}(q^2)$ and 
$\tilde{f}_{M0}(q^2)$ are finite at $q^2\to 0$ 
($q$ being the four-momentum transfer). 
The cLFV loops involving $h_1$ leptoquarks and up (or down) quarks
contribute to the off-shell penguin form factors as follows
\begin{align}\label{fu1} 
& f^{i=u}_{E0} \,=\,- f^{i=u}_{M0} \,=
-\;\frac{\sum_i X_{i\ell'}^*\,X_{i\ell}}{3\, (4\pi)^2}\,
\frac{(-q^2)}{m^2_{h_1}}\,
\left(\ln\frac{(-q^2)}{m^2_{h_1}} \,+ \,f_\gamma(r_i)\,-\,
\frac{1}{12}\right)\,, \nonumber \\
& f^{i=d}_{E0}\,=\,- f^{i=d}_{M0}\,=\,
-\;\frac{\sum_i  X_{i\ell'}^*\,X_{i\ell}}{6\, (4\pi)^2}\,
\frac{(-q^2)}{m^2_{h_1}}\,
\left(\ln\frac{(-q^2)}{m^2_{h_1}} \,+\,  f_\gamma(r_i)\,-\,
\frac{1}{3}\right)\, .
\end{align}
We recall that in the above     
$X_{i\ell}\equiv -\sqrt{2} y_{q\ell}$ for $i=d$, and $X_{i\ell}\equiv 
-V_{ij}^{*}y_{j\ell}$ for $i=u$; the loop function
$f_\gamma(r_i)$ is given by
\begin{equation}
f_\gamma(r_i)\,=\,-\frac{1}{3}\,+\,4 \,r_i\,+\,
\ln r_i \, +\,\left(1\,-\,2\,r_i \right)\,
\sqrt{1+4 r_i} \,\ln\left( 
\frac{\sqrt{1+4r_i}+1}{\sqrt{1+4r_i}-1}\right)\, .
\label{f1}
\end{equation}
with $r_i=m_{i}^2/(-q^2)$,  where $i$ denotes the quark in the loop. 

\noindent 
As an example, for the case of $\mu \to 3 e$ decays, one is led to the
following branching ratio~\cite{Okada:1999zk,Kuno:1999jp}
\begin{eqnarray}
\text{BR}(\mu \to e e e)&=&
2\,\left(|g_3|^2\,+\,|g_4|^2\right)
\,+\,|g_5|^2+|g_6|^2\,+\nonumber\\ 
&&+ 8\,e\, \text{Re}\left[C^{\mu e}_R\,
\left(2g_4^*\,+\,g_6^*\right)\,+\,C^{\mu e}_L \,
\left(2g_3^*\,+\,g_5^*\right)\right]\,+\nonumber\\ 
&&+ \frac{32\,e^2}{m_{\mu}^2}\,
\left\{\ln\frac{m_\mu^2}{m_e^2}\,-\,
\frac{11}{4}\right\}(\left|C_{R}^{\mu e}\right|^2\,+\,
\left|C_{L}^{\mu e}\right|^2)\,.
 \end{eqnarray}
Analogous expressions can be easily inferred for the other cLFV 3-body
decay channels.


\subsection{Neutrinoless $\mu$--$e$ conversion in Nuclei}
One of the most important constraints on SM extensions with scalar
leptoquarks\footnote{Recently, neutrinoless $\mu-e$ conversion in
  nuclei, in particular the comparative study of spin-independent
  versus spin-dependent contributions, 
  has been explored as a powerful means of disentangling distinct 
  leptoquark realisations~\cite{Davidson:2017nrp}.} 
arises from the nuclear assisted $\mu-e$ conversion. 
Phenomenologically - and contrary to other cLFV transitions which remain
loop-mediated processes - neutrinoless $\mu-e$ conversion can 
occur at the tree-level in the presence of lepton-quark-leptoquark
interactions. 
Moreover, as summarised in Table~\ref{tab:cLFV},
the experimental prospects for this cLFV
observable are particularly promising: not only current bounds
(obtained for Gold nuclei) are already $\mathcal{O}(10^{-13})$, but
in the near future several dedicated experiments should bring the
sensitivity down to
$\mathcal{O}(10^{-17,-18})$~\cite{Bartoszek:2014mya,Kuno:2005mm}.
The conversion ratio is defined as
\begin{equation}
\text{CR}(\mu-e,\text{N}) \, 
\equiv \frac{\Gamma(\mu-e,\text{N}) }{\Gamma_\text{capture}(Z)}
\end{equation}
in which $\Gamma_\text{capture}(Z)$ denotes the capture rate for a
nucleus with atomic number $Z$, and
$\Gamma(\mu-e,\text{N})$ is the cLFV width, which can 
be generically cast as follows~\cite{Kitano:2002mt,Dorsner:2016wpm}:
\begin{equation}\label{eq:conversion-rate}
\Gamma^{\mu-e} \,= \,2 \,G_F^2\, 
\left| \frac{C_{R}^{\mu e *}}{m_{\mu}}\, D \,+\,  
\left(2\,g_{LV}^{(u)} \,+\,g_{LV}^{(d)}\right) \,V^{(p)} 
\,+\, \left(g_{LV}^{(u)} \,+\, 2 \,g_{LV}^{(d)}\right)
\,V^{(n)}\right|^2 
\,+\, 2 \,G_F^2\left| \,\frac{C_{L}^{\mu e *}}{m_{\mu}}\, D \right|^2\,, 
\end{equation}
where the (tree-level) flavour violation is encoded
in the following quantities~\cite{Dorsner:2016wpm}
\begin{equation}
g_{LV}^{(d)} \,= \, -4 \frac{v^2}{2m_{h_1}^2} \, 
y_{d\ell'}\, y^{*}_{d\ell}\,,\quad \quad
g_{LV}^{(u)} \,= \,-4 \frac{v^2}{2m_{h_1}^2} \,  
(V^T\, y)_{u\ell'}\, (V^T \,y)^*_{u\ell}\,.
\end{equation}
Other than the (dominant) tree-level exchanges, 
we have also included
the photon-penguin contributions in the expression of the conversion
width\footnote{These are typically responsible for contributions to
  the conversion rate which are a factor of $10^{-3}$ smaller than the
tree-level contribution; box diagrams have also been found to provide
negligible contributions.}; 
these are associated with the 
$C_{L (R)}^{\ell\ell'}$ coefficients, which have been previously defined in 
Eq.~(\ref{eq:Cnorm}). 
The relevant nuclear information 
(nuclear form factors and averages over the atomic electric
field) are encoded in the 
$D$, $V^{(p)}$ and $V^{(n)}$ form factors. The latter overlap
integrals have been numerically estimated for various
nuclei~\cite{Kitano:2002mt}; Table~\ref{TabTiAuData} summarises some
of the above quantities for Gold and Aluminium nuclei 
(in units of $m_\mu^{5/2}$), as well as the corresponding 
capture widths.

Current bounds (from Gold nuclei) already allow to infer the following
stringent constraints~\cite{Dorsner:2016wpm}:
$g_{LV}^{(u)} < 8 \times 10^{-8}$ and $g_{LV}^{(d)} <
12 \times 10^{-8}$.

{\small
\renewcommand{\arraystretch}{1.2}
\begin{table}[t!]
\begin{center}
\begin{tabular}{l|ccc|c} \hline\hline
Nucleus & $D [m_\mu^{5/2}]$  & $V^{(p)} [m_\mu^{5/2}]$ &  
$V^{(n)} [m_\mu^{5/2}] $ & $\Gamma_{\text{capture}}[10^6 {s}^{-1}]$ 
\\ \hline\hline
$^{197}_{79}\text{Au}$ &  0.189 & 0.0974  & 0.146  &  13.07\\
$^{27}_{13}\text{Al}$  &  0.0362 & 0.0161 & 0.0173 &  0.7054\\
\hline\hline
\end{tabular}
\end{center}
\caption{Overlap integrals (normalised to units of $m_\mu^{5/2}$)
and total capture rates for Gold and Aluminium~\cite{Kitano:2002mt}.}
\label{TabTiAuData}
\end{table}
\renewcommand{\arraystretch}{1}
}

\subsection{Further leptonic observables}
Other tensions between SM predictions and observation have also
fuelled the need for NP. One such case is the anomalous magnetic
moment of the muon; we notice here that the present leptoquark
construction does provide a non-vanishing contribution to
$(g-2)_\mu$, albeit with the ``wrong'' sign~\cite{Dorsner:2017ufx}, so
that it cannot ease the current discrepancy.

Being a priori complex, the leptoquark couplings can also induce
contributions to the electric dipole moments of quarks and leptons, 
at the two loop level. Although these could possibly allow to further
constrain the couplings (in particular, the CP violating phases), such
a detailed analysis lies beyond the scope of the current
work.

\section{Accommodating $B$-anomalies, dark matter and neutrino
 data}\label{sec:results} 

In the previous sections we have discussed in detail the distinct
observations and experimental tensions that the present leptoquark
model is called upon to explain; moreover, we have also addressed a
comprehensive set of observables (encompassing numerous quark and
lepton flavour transitions) that are expected to lead to important
constraints on specific realisations of the model. 

In this section, we finally identify the different regimes thus allowing to:
\begin{enumerate}[(i)]
\item accommodate the latest data on neutrino oscillation parameters;
\item account for a correct relic abundance for the dark matter candidate;
\item explain the $R_{K^{(\ast)}}$ and $R_{D^{(\ast)}}$ anomalies; 
\item be compatible with all available bounds on leptoquark couplings
  and masses arising from direct searches, as well as from  
  the relevant leptonic and semi-leptonic meson decays and transitions
  (including neutral meson oscillations and rare meson decays) - both
  tree level and higher order processes -,   
  and cLFV processes (radiative and three-body decays, and $\mu-e$ conversion
  in nuclei).
\end{enumerate}

The results of the approximative numerical study of
Section~\ref{sec:dm} suggested that a viable dark matter candidate
could be obtained for a LZoP (the lightest $\Sigma^{1,0}$) mass in the
range $2.425~\text{TeV} \lesssim m_\Sigma  \lesssim 2.465~\text{TeV}$,
as inferred from Fig.~\ref{fig:dm}.
As working benchmark values, we will thus set the masses of the three
generations\footnote{Note that a priori there is no reason
   to assume a hierarchical structure for $m_{\Sigma}$; however, in
   the case of a degenerate spectrum, the Boltzmann equations relevant
   for calculating the dark matter relic abundance must take into
   account all three generations. Still, all 
   other (qualitative) conclusions would remain valid in such a case.}
of $m_{\Sigma^i}$ as 2.45, 3.5 and 4.5~TeV. 
By construction, the other $Z_2$-odd particle, $h_2$, must be necessarily
heavier than $\Sigma^1$; in order to comply with the hierarchy of the $Z_2$-odd
spectrum, we choose $m_{h_2}\sim 2.6$ TeV. 
Notice that $h_1$ is not subject to any DM-related arguments; its mass
is not related to that of the LZoP, nor to $m_{h_2}$, and 
can in principle vary in the TeV range. 

The chosen (illustrative) benchmark values of the scalar leptoquarks
and fermion triplets are in agreement with the current limits
established by negative collider searches; we
refer to~\cite{Dorsner:2017ufx,Cheung:2016frv} for 
a detailed discussion. We nevertheless highlight here a few important
points, and current experimental bounds. Both leptoquarks
can be pair produced via strong interactions
$pp\rightarrow h_{1(2)}h_{1(2)}$; each of the $Z_2$-even  
$h_1$ can subsequently decay into quark-lepton pairs 
(either neutrino or charged lepton). Searches for  
dilepton+dijet signals have been carried by the ATLAS and CMS
collaborations: for the 13~TeV run data, and considering decays into 
$ue$ and $c\mu$, ATLAS has set lower bounds on
leptoquark masses of $\gtrsim$ 1100~GeV (900~GeV), respectively
assuming 100\% (50\%) branching fractions~\cite{Aaboud:2016qeg};
mass limits on leptoquarks decaying to $b\tau$ have been established by
CMS, which has reported bounds $\gtrsim$ 850~GeV (550~GeV) assuming
100\% (50\%) branching fractions~\cite{Sirunyan:2017yrk}. 
The $Z_2$-odd $h_2$ can decay into $\Sigma$ and a down type quark; 
the decay mode associated with the neutral component of the triplet
leads to a dijet $+\slashed{E}_T$ signal (common to several
supersymmetry search channels). Preliminary bounds on the mass of $h_2$ can
thus be inferred from current squark mass limits:
about 1.3~TeV for first generation scalar down quarks~\cite{ATLAS:2016kts} 
and 800 GeV for third generation sbottoms~\cite{Aaboud:2016nwl}.

\bigskip
A survey of the previous sections dedicated to neutrino masses and
flavour observables reveals that the Yukawa couplings of the
leptoquark $h_1$ to matter 
are at the core of the distinct observables so far discussed: on the one
hand, $y$ is responsible for saturating the $B$-meson anomalies and
accounting for $\nu$-oscillation data (recall that $\tilde y$ can be
inferred from $y$ using a modified Casas-Ibarra parametrisation, see
Eq.~(\ref{eq:cip})); on the other, 
its different entries are severely  constrained by the strong bounds
arising either from negative searches or apparent SM-compatibility of
a vast array of flavour observables. 
Our first goal will thus be to identify the most minimal flavour
textures that can comply the with points listed above.

\subsection{Towards a parametrisation of the scalar leptoquark Yukawa 
couplings}\label{sec:texture}

Similarly to what occurs with the quark and lepton 
Yukawa couplings in the SM,
the new couplings are associated with numerous degrees of freedom
(being complex, $y$ contains 18 free parameters)
which, in the absence of a full theory of flavour, can only be
moderately constrained by data. 

A possible approach to circumvent the latter problem relies in
extending the symmetry group to include flavour symmetries, which can
effectively reduce the number of free parameters. 
To this end, there have been attempts to obtain hierarchical
leptoquark patterns by embedding the extended particle content in a
Froggatt-Nielsen framework, which can also explain the fermion mass
hierarchies as well as the CKM mixing pattern~\cite{Froggatt:1978nt}. 
The Froggatt-Nielsen (FN) mechanism is usually implemented via a U(1)
symmetry\footnote{Alternatively, a discrete $Z_N$ symmetry which
  becomes nearly continuous in the limit of large $N$, has also been
  considered, see for example~\cite{Hernandez:2016rbi,Hernandez:2015dga,Hernandez:2015hrt,Campos:2014zaa}.}
and a singlet scalar, non-trivially charged under the U(1)$_{\text{FN}}$. 
The singlet scalar then acquires a vacuum expectation value
$v_{_{\text{FN}}}$ at some high scale $\Lambda_{\text{FN}}$, resulting in a
suppression of the non-renormalisable Yukawa interactions by a factor
$({v_{_\text{FN}}}/{\Lambda_\text{FN}})^n$, 
where $n$ is the sum of the
fermion U(1)$_{\text{FN}}$ charges~\cite{Deppisch:2016qqd}. 
Alternatively, a weakly broken U(2)$^5$ flavour symmetry
has also been proposed in the context of possible interpretations of the
$B$-decay anomalies~\cite{Barbieri:2015yvd}. 

A systematic and comprehensive study of the allowed textures for the
leptoquark couplings, relying on symmetry-inspired flavour
constructions, clearly lies beyond the scope of the analysis; 
here, we will adopt a phenomenological approach, and identify possible
textures for the new Yukawa couplings from the requirements of
explaining the $B$-meson anomalies while complying with all available
experimental bounds.
As a starting point (inspired by generic FN-like flavour patterns), 
we consider generic parametrisations of $y$ in terms of powers of
a small parameter $\epsilon$ (taken to be positive and real), with each
entry\footnote{Notice that the Yukawa matrix $y$ is a priori 
a complex matrix in flavour space; however, for simplicity, we will
only consider real values both for $\epsilon$ and for $a_{ij}$.} 
weighed by an $\mathcal{O}(1)$ real coefficient $a_{ij}$:
\begin{equation}\label{eq:ytexture.1}
y_{ij} \,=\, a_{ij} \phantom{|}_\odot \,\epsilon^{n_{ij}}\,,
\end{equation}
with $\odot$ denoting that there is no summation implied over $i,j$. 

As a first step, we set the individual coefficients $a_{ij}=1$, and
use the requirement of saturating the $R_{K^{(*)}}$ tensions to infer
the size of the parameter $\epsilon$:
in Section~\ref{sec:meson} we have seen that at the leading order,
the explanation of the $R_{K^{(*)}}$ anomalies constrains combinations
of the quark-``muon'' couplings $y_{22}$ and $y_{32}$ (further
depending on inverse powers of the $h_1$ leptoquark mass). 
For a benchmark value of $m_{h_1}\sim 1.5$~TeV, one is led to the
following relation 
\begin{align}\label{eq:ytexture.2}
y_{22} \, y_{32}\, \approx \, 2.1555 \times 10^{-3} \, 
\sim  \, \epsilon^{n_{22}+n_{32}} \, \quad
\rightarrow \quad \, 
\epsilon^4 \, \sim \,2.1555 \times 10^{-3}\,
\Leftrightarrow \, \epsilon\, \approx \, 0.215 \,,
\end{align}
in which we have elected $n_{22}+n_{32}=4$ as a 
natural choice (so that $\epsilon \sim \mathcal{O}(1)$).

Following the above, and having fixed 
$\epsilon \approx 0.215$ (notice that this value reflects the
choice of $m_{h_1}$), we express the most general
texture written in terms of the parameter $\epsilon$ and
positive integers $n_{ij}\geq 1$ with $i,j=1,2,3$:  
\begin{align}
 y \,\sim \,\left( 
\begin{array}{ccc}
\epsilon^{n_{11}} & \epsilon^{n_{12}} & \epsilon^{n_{13}} \\ 
\epsilon^{n_{21}} & \epsilon^{n_{22}} & \epsilon^{n_{23}} \\ 
\epsilon^{n_{31}} & \epsilon^{n_{32}} & \epsilon^{n_{33}}
\end{array}
\right)\,, \label{eq:gentexture}
\end{align}
subject to the constraint $n_{22}+n_{32}=4$ to explain the
$R_{K^{(*)}}$ anomalies.
The experimental bounds on rare
meson and charged lepton decays can now
be used to identify generic textures\footnote{Another approach to
  constrain $y$ would be to consider minimal textures exhibiting
  vanishing entries (``texture zeroes'') in a given weak basis; 
  however, in the absence of an underlying symmetry, we prefer to
  consider the most general pattern for $y_{ij}$.} 
for $y$ which are compatible with observation.  

As can be inferred from the analytical expressions presented in 
Sections~\ref{sec:mesoncon} and~\ref{sec:lfv}, the Yukawa couplings of
$h_1$ to the first two generations of quarks are stringently
constrained from rare meson decays; likewise, its couplings to the
first generations of leptons are expected to be limited by
cLFV transitions in the $\mu-e$ sector. 
A numerical scan of all possible textures (i.e., thorough 
tests on the viability of
each $y(n_{ij})$ - cf. Eq.~(\ref{eq:gentexture}), 
for fixed values of $\epsilon$ and $m_{h_1}$ and with
$n_{22}+n_{32}=4$) has shown that the most constraining observables
turn out to be the rare decay $K^{+}\to \pi^{+} \nu \bar{\nu}$ and, on
the lepton sector, $\mu-e$ conversion in nuclei and the radiative
$\mu\to e\gamma$ decay.

The numerical study has further allowed to identify generic classes of 
representative textures which are in agreement with the $B$-meson
anomalies as well as all leptonic and mesonic processes taken into
account (for the above mentioned $(\epsilon,m_{h_1})$ benchmark): 
$\mu\to e \gamma$, $\tau\to e \gamma$, $\tau\to \mu \gamma$, 
$\mu \to 3 e$, $\tau \to 3 \mu$, $\mu- e$ conversion, 
and $K^{+}(K_L)\to \pi \nu \bar{\nu}$, 
$B\to K^{\ast} \nu \bar{\nu}$, $B_s^{0}-\bar{B}_{s}^{0}$ 
oscillations\footnote{In the subsequent discussion we do
  not include constraints from $K^0-\bar K^0$ mixing, as the latter
  was found to provide weaker constraints than those arising from 
  $K\to \pi \nu \bar{\nu}$ decays.}, 
as well as the cLFV decays $B_s \to \mu e$ and $K_L \to \mu e$.  

The three classes of textures are identified by the specific
realisation of $(n_{22},n_{32})$: $(3,1)$ - type I; 
$(2,2)$ - type II; $(1,3)$ - type III. For each class, the allowed
textures are presented in Table~\ref{table:textures:general}.
{\small
\renewcommand{\arraystretch}{1.2}
\begin{table}[h!]
\begin{center}
\begin{tabular}{|c|c|c|c|}
\hline
& Type I & Type II  & Type III  \\
\hline
$y$ & 
$\begin{pmatrix}
\times &\times & \times \cr 
\times & \epsilon^{3} & \times \cr
\times & \epsilon & \times
\end{pmatrix}^{\phantom{|}}_{\phantom{|}}$  &
$\begin{pmatrix}
\times &\times & \times \cr 
\times & \epsilon^{2} & \times \cr
\times & \epsilon^{2} & \times
\end{pmatrix}$  & 
$\begin{pmatrix}
\times &\times & \times \cr 
\times & \epsilon & \times \cr
\times & \epsilon^{3} & \times
\end{pmatrix}$
\\
\hline
$\begin{matrix} \text{Generic allowed}\cr
                \text{textures} \end{matrix}$	&  
$\begin{pmatrix}
\epsilon^{4} &\epsilon^{\geq 5} & \epsilon^{\geq 2} \cr 
\epsilon^{\geq 3} & \epsilon^{3} & \epsilon^{\geq 4} \cr
\epsilon^{\geq 4} & \epsilon & \epsilon^{\geq 1}
\end{pmatrix}^{\phantom{|}}_{\phantom{|}}$  &
$\begin{pmatrix}
\epsilon^{6} &\epsilon^{\geq4} & \epsilon^{\geq 3} \cr 
\epsilon^{\geq 5} & \epsilon^{2} & \epsilon^{\geq 3} \cr
\epsilon^{\geq3} & \epsilon^{2} & \epsilon^{\geq 1}
\end{pmatrix}$  & 
$\begin{pmatrix}
\epsilon^{5} &\epsilon^{\geq5} & \epsilon^{\geq 4} \cr 
\epsilon^{4} & \epsilon & \epsilon^{\geq 2} \cr
\epsilon^{\geq4} & \epsilon^{3} & \epsilon^{\geq 1}
\end{pmatrix}$ \\
\hline
\end{tabular}
\caption{Classes of textures for the $y$ couplings complying
  with the constraint $n_{22}+n_{32}=4$: type I, II and III.
  For a small parameter ($\epsilon \sim 0.215$), the second row displays
  the generic allowed textures in terms of powers of $\epsilon$, 
  consistent with the current experimental bounds on the leptonic
  processes $\ell \to \ell^\prime \gamma$, $\ell \to 3\ell^\prime$, 
  $\mu -e$ conversion in nuclei, 
  $K^{+}(K_L)\to \pi^{+} (\pi^{0}) \nu \bar{\nu}$, $B\to
  K^{\ast} \nu \bar{\nu}$, $B_s^{0}-\bar{B}_{s}^{0}$ oscillation and $B_s
  \to \mu e$, $K_L \to \mu e$. 
}\label{table:textures:general}
\end{center}
\end{table}
\renewcommand{\arraystretch}{1.}
}

For each of the classes identified, we have chosen an illustrative
case (setting $n_{ij}$ in agreement with
Table~\ref{table:textures:general}), and 
we have evaluated the associated
contributions to the different leptonic and mesonic observables
mentioned above. The information is summarised in  
Table~\ref{tab:texture}. 
We do not include here 
bounds from the neutral meson-antimeson oscillations 
since they are considerably less constraining than 
the meson decay processes involving the same set of leptoquark couplings.
{\small
\renewcommand{\arraystretch}{1.2}
\begin{table}[h!]
\begin{center}
\begin{tabular}{|c|c|c|c|}
\hline
& Type I & Type II  & Type III \\
\hline
$y$ example & $\begin{pmatrix}
\epsilon^{4} &\epsilon^{5} & \epsilon^{2} \cr 
\epsilon^{3} & \epsilon^{3} & \epsilon^{ 4} \cr
\epsilon^{4} & \epsilon & \epsilon
\end{pmatrix}^{\phantom{|}}_{\phantom{|}}$  &
$\begin{pmatrix}
\epsilon^{6} &\epsilon^{4} & \epsilon^{3} \cr 
\epsilon^{ 5} & \epsilon^{2} & \epsilon^{ 3} \cr
\epsilon^{3} & \epsilon^{2} & \epsilon
\end{pmatrix}$  & 
$\begin{pmatrix}
\epsilon^{5} &\epsilon^{5} & \epsilon^{4} \cr 
\epsilon^{4} & \epsilon & \epsilon^{2} \cr
\epsilon^{4} & \epsilon^{3} & \epsilon
\end{pmatrix}$ \\
\hline\hline
$\text{BR}(\mu\to e \gamma)$ & $1.21 \times 10^{-13}$  & 
$8.99 \times 10^{-14}$ & $8.31 \times 10^{-14}$ \\
\hline
$\text{BR}(\tau \to \mu \gamma)$ & $1.47\times 10^{-10}$  & 
$7.45 \times 10^{-12}$ & $9.46 \times 10^{-12}$ \\
\hline
$\text{BR}(\tau \to e \gamma)$ & $2.31 \times 10^{-14}$  & 
$3.17 \times 10^{-13}$ & $2.14 \times 10^{-14}$ \\
\hline
$\text{BR}(\mu \to 3 e )$ & $2.73 \times 10^{-13}$ & 
$2.02 \times 10^{-13}$ & $2.02 \times 10^{-13}$ \\
\hline
$\text{BR}(\tau \to 3 \mu )$ & $1.92 \times 10^{-9}$ & 
$9.49 \times 10^{-11}$ & $1.30 \times 10^{-10}$ \\
\hline
$\text{BR}(\tau \to 3 e )$ & $2.93 \times 10^{-13}$ & 
$4.01 \times 10^{-12}$ & $2.73 \times 10^{-13}$ \\
\hline
$\text{CR}(\mu - e, \text{ N})$ & $1.81 \times 10^{-13}$ & 
$2.75 \times 10^{-14}$ & $1.61\times 10^{-13}$\\
\hline	\hline
$\text{BR}(K^+ \to \pi^+ \nu\bar \nu)$ & $1.32 \times 10^{-10}$ & 
$1.21 \times 10^{-10}$ & $1.22 \times 10^{-10}$\\
\hline
$\text{BR}(K_L \to \pi^0 \nu\bar \nu)$ & $3.25 \times 10^{-11}$ & 
$3.10 \times 10^{-11}$ & $3.10 \times 10^{-11}$\\
\hline	
$R^{\nu\nu}_{K^{(\ast)}_{\phantom{i}}}$ & $1.04$ & $1.08$ & $1.53$\\
\hline	\hline
$\text{BR}(K_L \to \mu e)$ & $1.96 \times 10^{-13}$ & 
$9.06 \times10^{-15}$ & $9.06 \times10^{-15}$\\
\hline
$\text{BR}(B_s \to \mu e)$ & $3.13 \times 10^{-15}$ & 
$1.47\times 10^{-12}$ & $1.47\times 10^{-12}$\\
\hline	\hline
\end{tabular}
\caption{Contributions to distict observables associated with
  illustrative examples of each of the texture classes given in
  Table~\ref{table:textures:general} (viable for the 
  benchmark choice $(\epsilon,m_{h_1})=(0.215, 1.5~\text{TeV})$). 
}\label{tab:texture}
\end{center}
\end{table}
\renewcommand{\arraystretch}{1.}
}

\begin{figure}[h!]
\begin{center}
\includegraphics[width=0.8\textwidth]{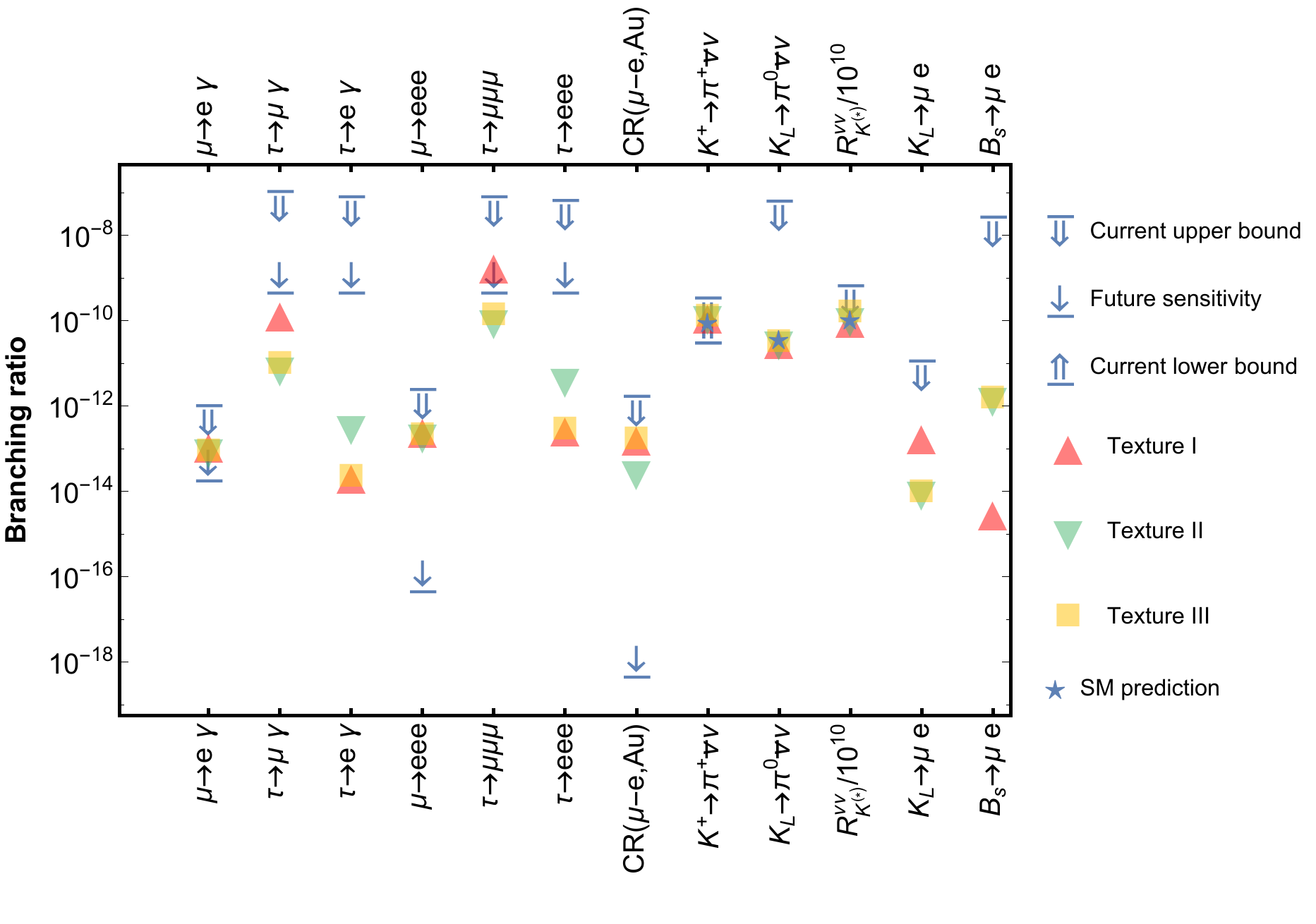}
\caption{Contributions to several leptonic and mesonic observables
  associated with the three textures I, II, and III for the 
  benchmark choice $(\epsilon,m_{h_1})=(0.215, 1.5~\text{TeV})$: 
  $\mu\to e \gamma$, $\tau\to \mu \gamma$, $\tau\to e \gamma$, 
  $\mu \to 3 e$, $\tau \to 3 \mu$, $\mu - e$ conversion, 
  $K^{+}(K_L)\to \pi^{+} (\pi^{0}) \nu \bar{\nu}$, 
  $R^{\nu\nu}_{K^{(\ast)}} /10^{10}$,
  $B_s \to \mu e$, and $K_L \to \mu e$. 
  The relevant experimental bounds (and future sensitivities),
  as well as SM predictions (when appropriate), are also displayed.} 
\label{fig:summary}
\end{center}
\end{figure}

Figure~\ref{fig:summary} graphically summarises the information given
in Table~\ref{tab:texture}: for each type of texture, we display the
associated predictions for $\mu\to e \gamma$, $\tau\to \mu \gamma$, 
$\tau\to e \gamma$, $\mu \to 3 e$, $\tau \to 3 \mu$, 
$\mu - e$ conversion, $K^{+}(K_L)\to \pi^{+} (\pi^{0}) \nu \bar{\nu}$, 
$R^{\nu\nu}_{K^{(\ast)}}$ as well as 
$B_s \to \mu e$ and $K_L \to \mu e$. 
For each process we include the current experimental bounds and future
sensitivities, and when applicable, the SM predictions.

\bigskip
Before proceeding, we briefly comment on the other LFUV observables
discussed in Section~\ref{sec:RD*}.
The present leptoquark construction leads to SM-like predictions to 
the distinct $R_{D^{(\ast)}}$ ratios (independently of the texture
type and/or mass regime for the exchanged pseudoscalar). 
If on the one hand this means that, once the distinct experimental
constraints have been taken into account, the muon to electron ratios 
$R_{D^{(\ast)}}^{\mu/e}$ are consistent with experimental
measurements, on the other hand it also implies that the current
experimental measurement of $R_{D^{(\ast)}}$ (tau to muon ratio,
exhibiting a significant deviation from SM predictions) cannot
be accounted for. Should the latter $R_{D^{(\ast)}}$ discrepancy
be confirmed in the future, then the
present leptoquark construction will be ruled out, at least in this
minimal version.

\subsection{Constraining the leptoquark parameter space}\label{sec:results2}

The previously 
chosen textures, as well as the numerical results (both for the
$\epsilon$ parameter and for the contributions to the
distinct observables) were obtained for a benchmark value of the
$h_1$ leptoquark mass, $m_{h_1}=1.5$~TeV. The natural question to
address is how the viability of the model is impacted by different
choices of its parameters, in particular the entries of the $y$
couplings and $m_{h_1}$. 

To explain the
$R_{K^{(*)}}$ anomalies, the BSM construction must 
comply with the conditions given in
Section~\ref{subsec:RK}, in particular with the interval for the 
$C^{ee,\mu\mu}_{9,\text{NP}}$ couplings given in
Eq.~(\ref{eq:mu.e.fit}); 
this can also be written as a condition on the ratio of the relevant
Yukawa couplings to the $h_1$ leptoquark mass, 
$0.64 \times 10^{-3}\lesssim \tfrac{{\rm\, Re}[y_{b\mu}y_{s\mu}^\ast -
y_{be}y_{se}^\ast]}{(m_{h_1}/1 {\rm ~TeV})^2 } \lesssim 1.12 \times 10^{-3}$.

\begin{figure}[h!]
\begin{center}
\mbox{\includegraphics[width=0.49\textwidth]{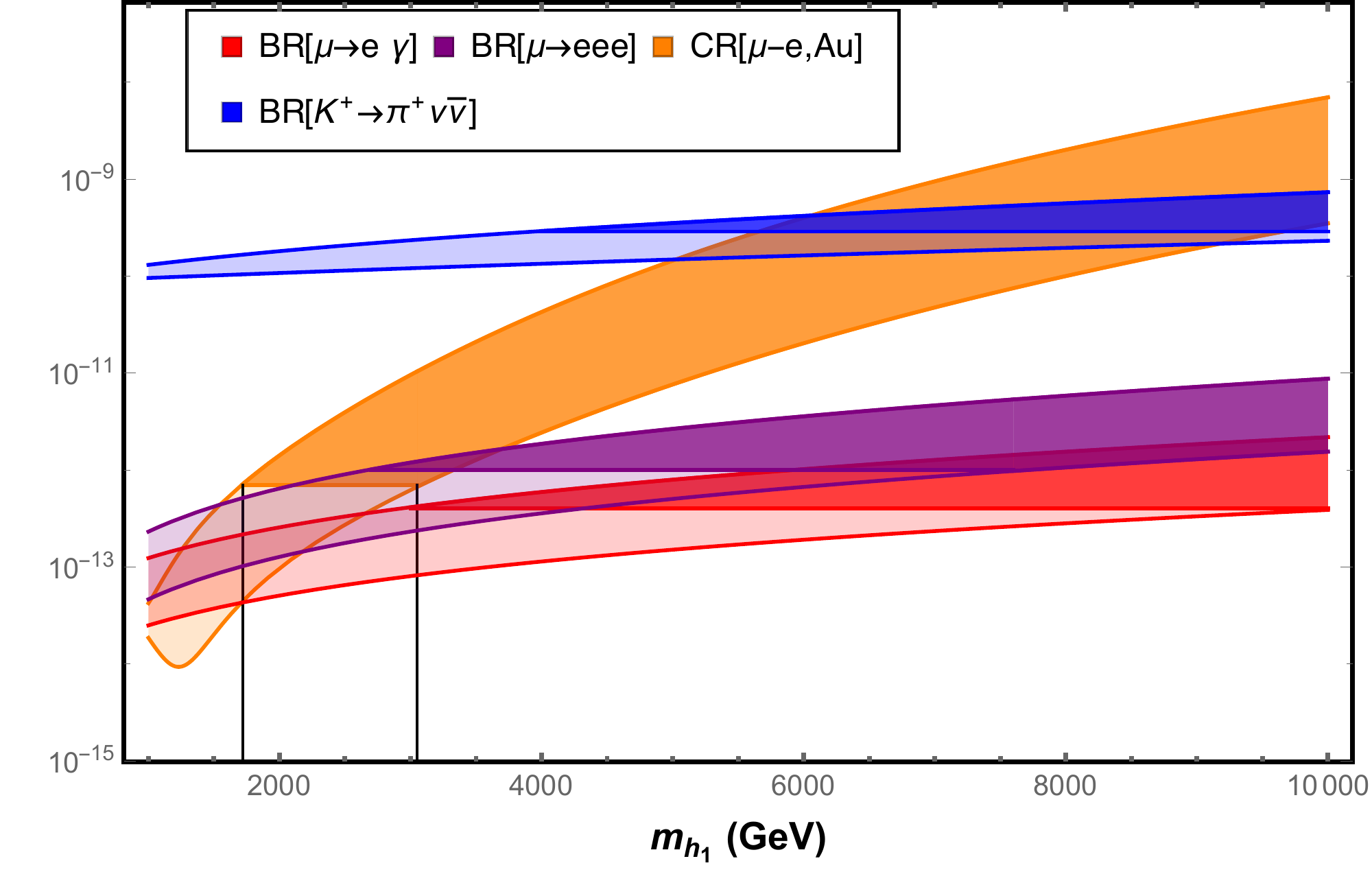}
\hspace*{5mm}
\includegraphics[width=0.49\textwidth]{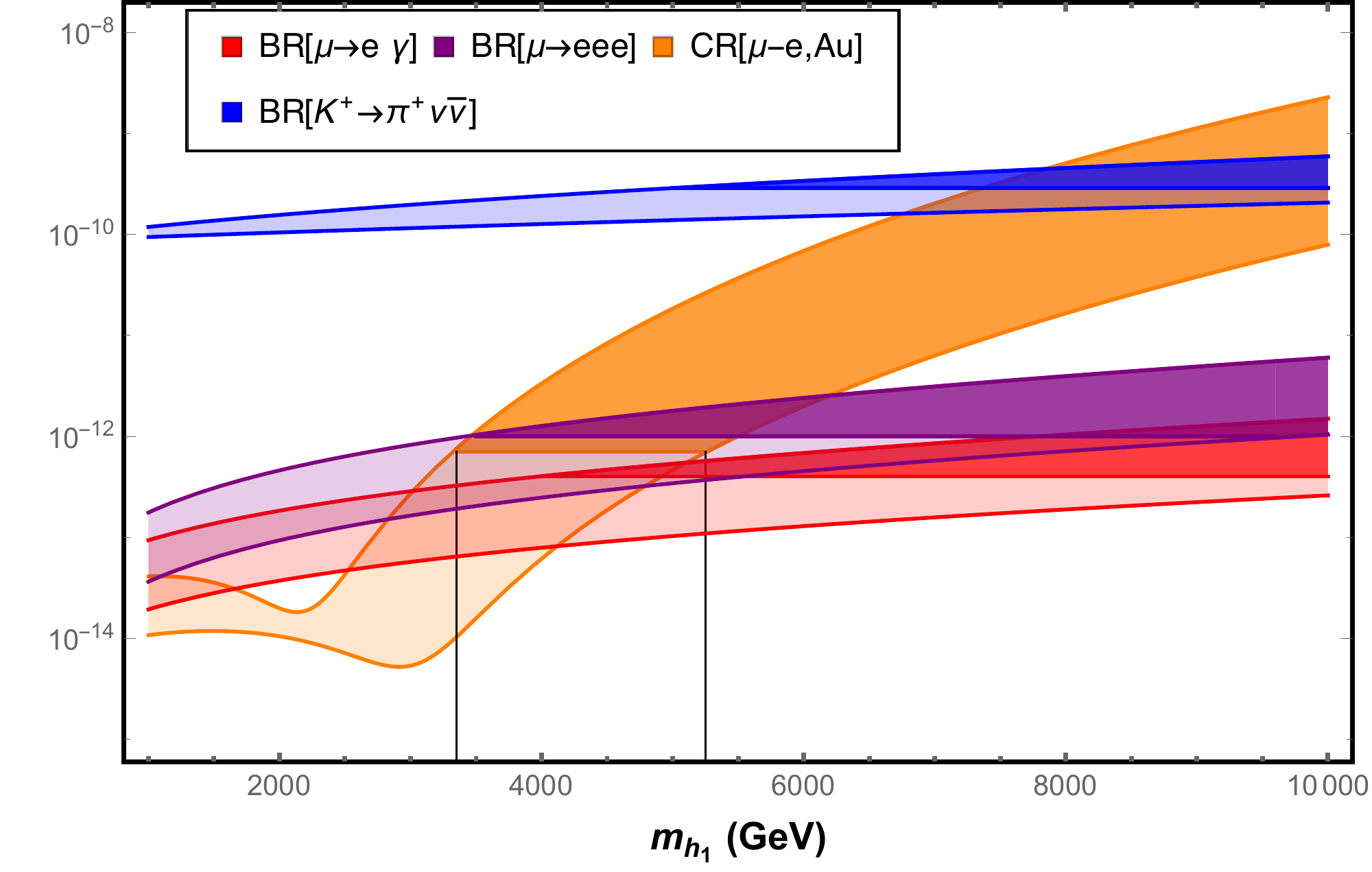}}\\
\vspace*{3mm}
\includegraphics[width=0.49\textwidth]{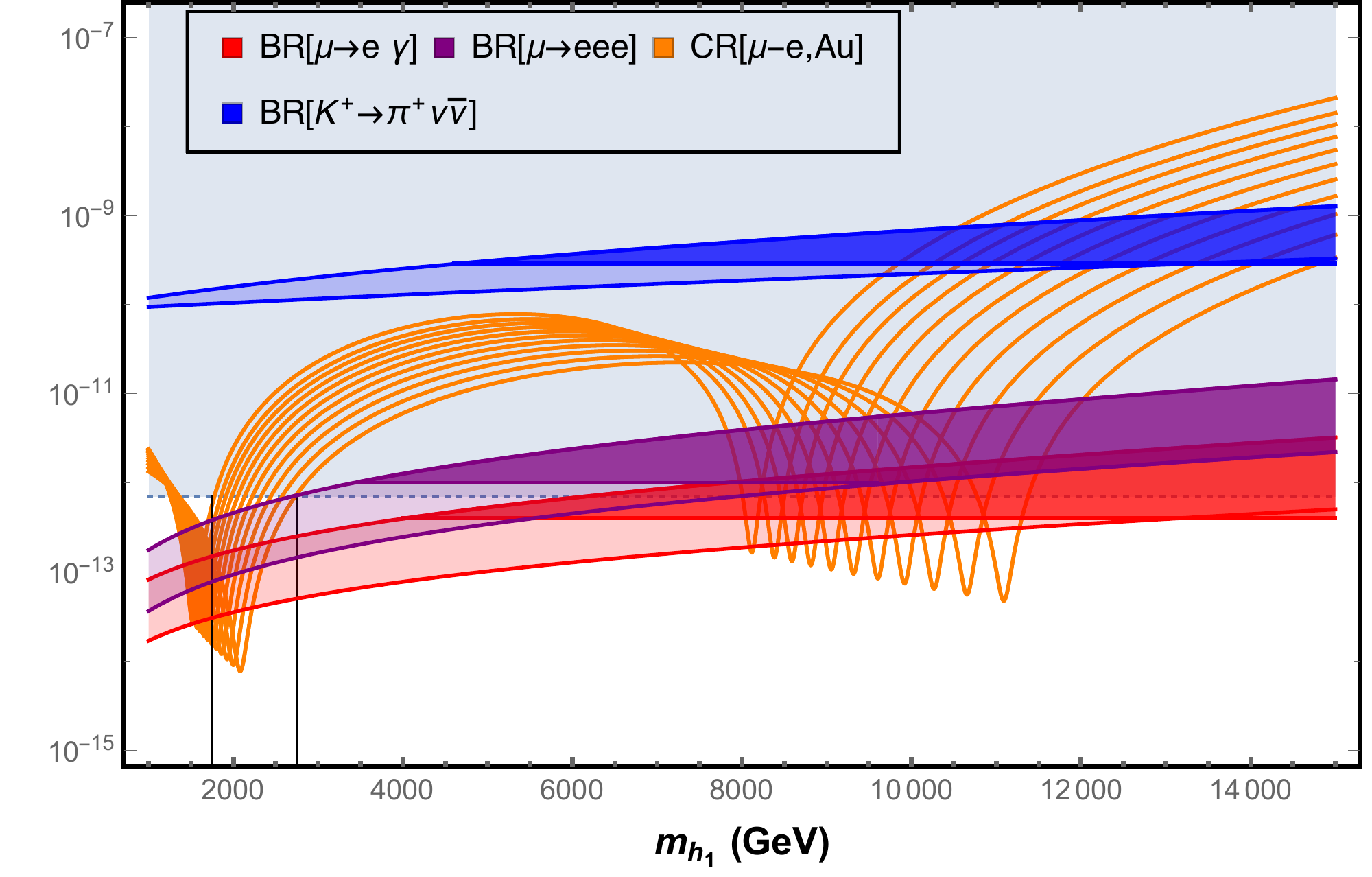}
\caption{Contributions to BR($\mu \to e \gamma$), 
BR($\mu \to 3e$), CR($\mu -e$, Au) and 
BR($K^+ \to \pi^+ \nu \bar \nu$) 
as a function of the $h_1$ leptoquark
mass $m_{h_1}$, for type I, II and III textures 
of $y$ (respectively from left to right, top to bottom),
complying with the current interval for $R_{K^{(*)}}$. 
Light (solid) surfaces denote currently allowed (excluded) regimes due
to the violation of the associated experimental bound.} 
\label{fig:textures.mh1}
\end{center}
\end{figure}

Varying the mass of the $h_1$ leptoquark over a wide interval - in
agreement with LHC direct search bounds - leads to new ranges for the
relevant entries of the Yukawa couplings $y_{ij}$ (and thus new values
for $\epsilon$). Since for increasing values of $m_{h_1}$ saturating
the $R_{K^{(*)}}$ anomalies calls for larger $y_{22,32}$ - and hence
for larger $\epsilon$ -, bounds on other observables are expected to
become more severe, leading to the exclusion of a given realisation.
This is displayed on the distinct panels of
Fig.~\ref{fig:textures.mh1}: the coloured regions denote the
contributions for a given observable arising from varying $\epsilon$
(i.e., $y_{ij}$) 
in the $R_{K^{(*)}}$ favoured interval given above. Light (dark)
regions correspond to allowed (excluded) regimes in view of current
experimental bounds. 

Leading to Fig.~\ref{fig:textures.mh1} we have elected to plot only
the most constraining observables: BR($\mu \to e \gamma$), 
BR($\mu \to 3e$), CR($\mu -e$, Au) and 
BR($K^+ \to \pi^+ \nu \bar \nu$). 
Concerning textures of type I and II, one can verify that 
CR($\mu -e$, Au) precludes values of the leptoquark mass 
respectively larger than 
$m_{h_1} \gtrsim 1.8$~TeV and 3.4~TeV (for $\epsilon_\text{max}$)
and 
$m_{h_1} \gtrsim 3$~TeV and 4.2~TeV (for $\epsilon_\text{min}$).
(Notice that for texture II $\mu \to 3e$ is almost as constraining as
$\mu-e$ conversion in nuclei.) 
For type III textures, one finds two intervals for $m_{h_1}$: 
[1.75~TeV, 2.75~TeV] and [8~TeV, 11.4~TeV]
(the lower and upper bounds obtained in
association with the maximal and minimal values of $\epsilon$).

Despite being manifest in all panels, the ``kinks'' associated with the
contributions to CR($\mu -e$, Au) are particularly apparent for type
III textures\footnote{For this reason we have preferred to display
  several lines associated with values of $\epsilon$, which is varied with
a $5 \times 10^{-5}$ step.}. The behaviour has been well identified
in the literature (see, e.g.~\cite{Alonso:2012ji}), and stems from
having a localised cancellation of opposite sign 
up- and down-type quark contributions to the conversion rate 
(due to different charge and weak isospin). 

\bigskip
At this point, it is important to recall that the proposed
parametrisation for $y$, as given in Eq.~(\ref{eq:ytexture.1}),
allows each element to be weighed by a real coefficient $a_{ij}$, 
$\mathcal{O}(1)$. 
In order to understand how generic perturbations of the unconstrained
entries of $y$ affect the phenomenological viability of the model, 
we have thus taken a type I texture for $y$, and 
varied one $a_{ij}$ at a time\footnote{We fix $y_{22,32}$
  to ensure that $R_{K^{(*)}}$ remains accounted for, hence
  $a_{22,32}=1$.} in the range [0.4,~1.6] (with a step size of 0.1).  
Although not explicitly displayed here,  
the numerical studies revealed that 
$\mu \to e \gamma$ and $\mu \to 3 e$ are predominantly sensitive to 
$y_{31}$ ($a_{31}$), with a mild secondary dependence on $y_{21}$
($a_{21}$); likewise,  
$\tau \to \mu \gamma$ and $\tau \to 3 \mu$ are controlled by 
$y_{33}$ ($a_{33}$); $\tau \to e \gamma$ and $\tau \to 3 e$ are
sensitive to variations from both $y_{31}$ and $y_{33}$.
Finally, $K^+ \to \pi^+ \nu
\bar \nu$ exhibits a significant dependence on $y_{13}$ ($a_{13}$), 
$y_{21}$ ($a_{21}$) and $y_{23}$ ($a_{23}$).
As a general qualitative statement, for all the radiative and 3-body
decays mentioned above, the variation of the $a_{ij}$ coefficients in
the interval [0.4,~1.6] leads to a variation of about one order of
magnitude in the prediction for the observable. 
Occurring at the tree level, the neutrinoless $\mu-e$ conversion
strongly depends on $y_{11}$ ($a_{11}$), $y_{12}$
($a_{12}$), and $y_{21}$ ($a_{21}$) - the dominant element depending
to a certain extent on the leptoquark mass regime.

The same study can be carried for type II and III textures, with
the results reflecting the relative $\epsilon^{n_{ij}}$ dependence.

\subsection{Final constraints from neutrino oscillation data}
The requirements of having a viable DM candidate, and of accounting for the 
$R_K^{(*)}$ anomalies while complying with all available data on meson
and lepton rare decays and transitions have allowed to identify viable
flavour textures for the $y$ leptoquark Yukawa couplings, as well as 
mass regimes for the new states. 

As mentioned in Section~\ref{sec:neutrino},
once the flavour structure of $y$ has been fixed (be it from
theoretical arguments or, as in the present case, from a comprehensive
phenomenological analysis), the modified Casas-Ibarra parametrisation
of Eq.~(\ref{eq:cip})
readily allows to determine $\tilde y$, while complying with current
neutrino oscillation data. 
As a final step in our study, we thus consider the three textures
already discussed in the previous section, and for each one we vary the
$n_{ij}$ powers of $\epsilon$ in agreement with the ranges given in
Table~\ref{table:textures:general}, 
as well as the associated $a_{ij}$ prefactors (in the
range [0.4,1.6]). The $h_1$ leptoquark mass is, for
simplicity, set to the benchmark value of 1.5~TeV (although the results
here discussed qualitatively hold for other choices - in agreement
with the discussion of the previous subsection).

Concerning neutrino data, we use the best-fit values 
from the global oscillation analysis of~\cite{deSalas:2017kay}, 
taking a normal ordering for the neutrino spectrum, 
with the lightest neutrino mass taken in the range
$[10^{-8}~\text{eV}, 0.001~\text{eV}]$.
As already mentioned, we take the
right-handed triplet masses to be $m_{\Sigma}$=2.45, 3.5 and 4.5~TeV.
The remaining degrees of freedom in the modified Casas-Ibarra
parametrisation are randomly sampled from the following intervals:
$[0,2\pi]$ for the phases, and $[-4\pi,4\pi]$ for the angles; one
further has $\lambda_h \lesssim 4 \pi$ (see Eq.~(\ref{lag3.1})).

Each of the thus obtained couplings ($y$ and $\tilde y$) are again
subject to the various flavour constraints 
previously discussed; moreover, each
entry of the couplings must comply with perturbativity requirements, 
$|{y} (\tilde y)| \lesssim 4\pi$.

\begin{figure}[h!]
\begin{center}
\includegraphics[width=0.5\textwidth]{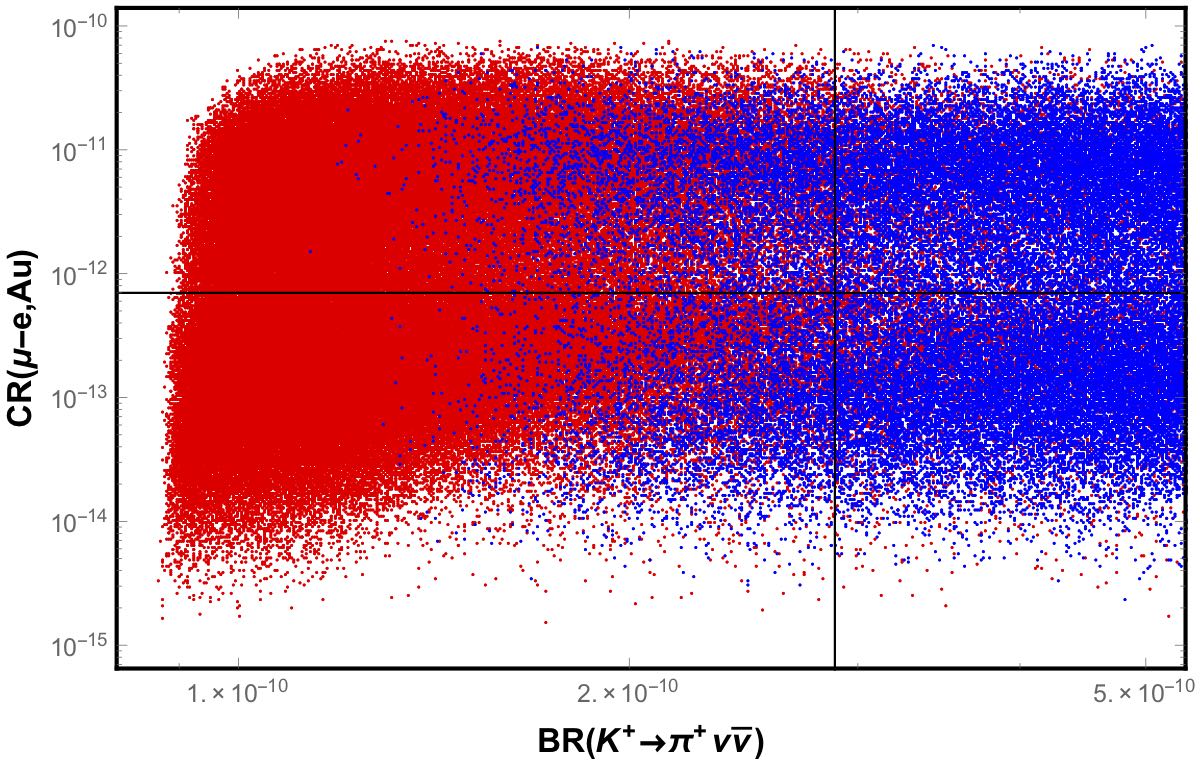}
\vspace*{3mm}\\
\hspace*{-3mm}\mbox{\includegraphics[width=0.50\textwidth]{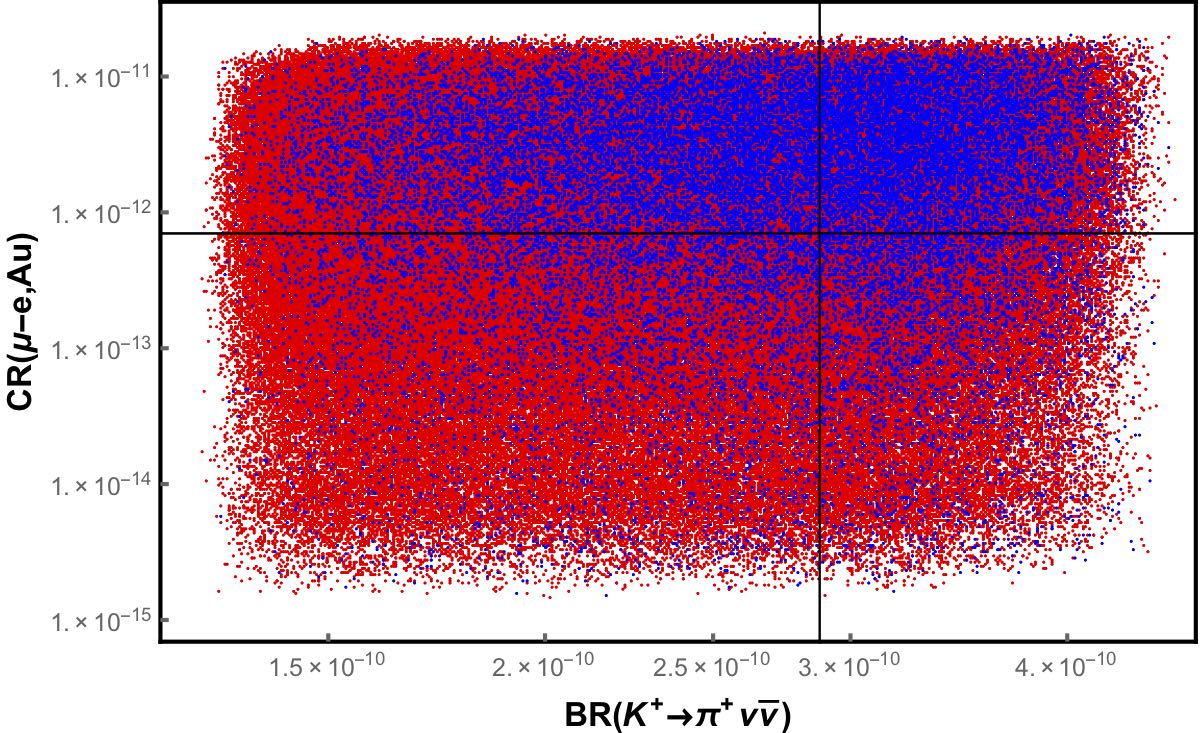}
\hspace*{3mm}
\includegraphics[width=0.50\textwidth]{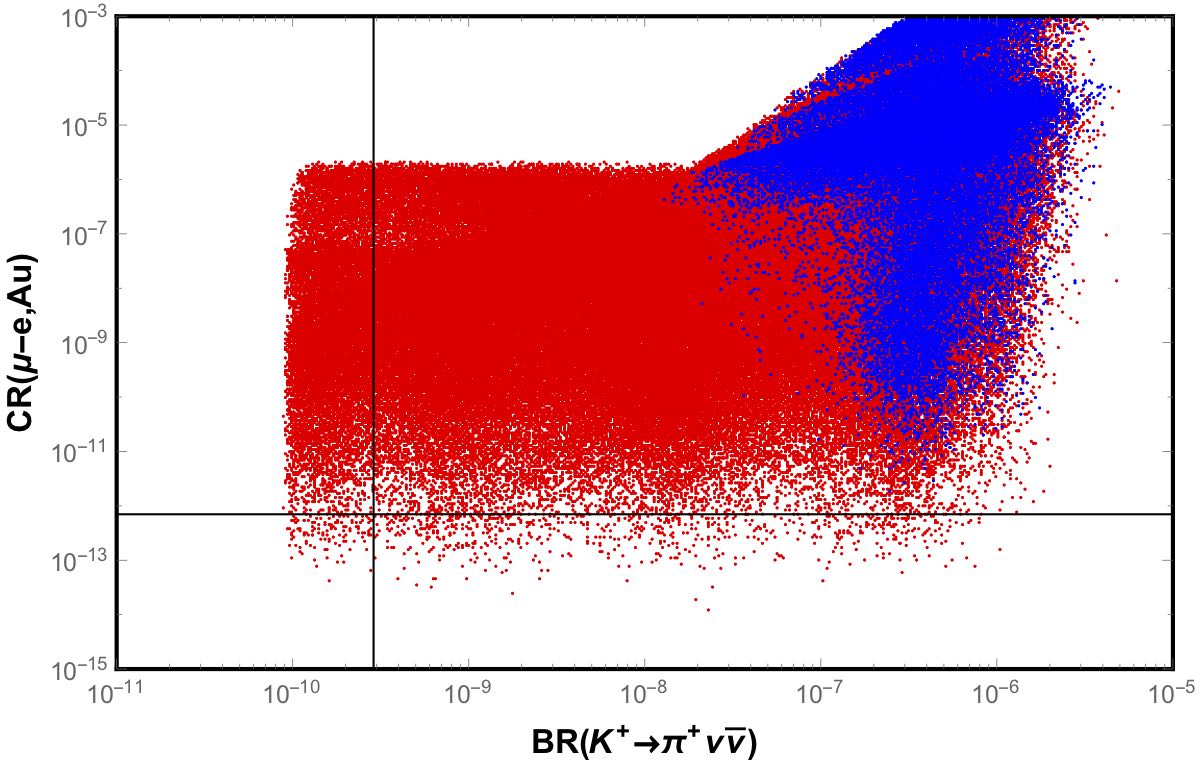}}
\caption{Predictions for CR($\mu-e$, N) and 
BR($K^+ \to \pi^+ \nu \bar \nu$)
associated with randomly sampled $y$ and $\tilde y$ (see text), for
type I, II and III textures (top to bottom, left to right). 
The horizontal/vertical lines denote the current experimental bounds,
and the colour code identifies perturbative (blue) or non-perturbative 
(red) regimes of $\tilde y$. } \label{fig:CI:results}
\end{center}
\end{figure}

In order to illustrate our findings, we display in
Fig.~\ref{fig:CI:results} the results of the scan, for the three types
of textures. Since neutrinoless conversion in nuclei and the 
$K^+ \to \pi^+ \nu \bar \nu$ decay are the most constraining
observables, we display the corresponding predictions of the
randomly sampled textures in the plane spanned by the latter two
observables; the colour code distinguishes between perturbative and
non-perturbative entries of the $y$ and $\tilde y$ couplings. 

As is manifest from inspection of Fig.~\ref{fig:CI:results}, 
accommodating $\nu$-oscillation data from  type I textures for the
leptoquark $y$ couplings, in agreement with experimental data,
and for perturbative $\tilde y$ does not excessively constrain the
remaining degrees of freedom. 
Even though perturbative regimes for $\tilde y$ are more likely to be
associated with large values of  BR($K^+ \to \pi^+ \nu \bar \nu$), 
one can easily find regimes which are phenomenologically allowed.
Notice however that a near-future improvement in the associated
experimental sensitivities (for instance CR($\mu-e$, Al)$~\sim
10^{-15}$, and  BR($K^+ \to \pi^+ \nu \bar \nu$)$\sim 10^{-10}$)
should allow to probe the present leptoquark construction (and
possibly falsify it). 

For type II textures, the associated panel of
Fig.~\ref{fig:CI:results} reveals that perturbative $\tilde y$
couplings are far harder to accommodate, especially due to the
excessive contributions to CR($\mu-e$, Au). Finally, notice that only
a tiny subset of the sampled type III textures is in
agreement with flavour observables, and no sub-region of the latter 
leads to perturbative $\tilde y$. 
One thus concludes that type III textures (despite being marginally
compatible with all the quark and lepton observables here discussed)
do not lead to a satisfactory leptoquark construction. 

We have also explored the possibility of having a distinct ordering
(inverted) for the light neutrino spectrum: in what concerns type I
textures, we found no significant changes, so that in fact both
orderings can be easily accommodated; in the case of 
type II textures for $y$ (which
do allow to accommodate oscillation data for a normal ordering) we
failed to find viable solutions for an inverted ordering; 
finally, type III textures remain unable to account for oscillation
data with perturbative values of the couplings even in the case of a 
inverted ordering of the neutrino spectrum.

Before moving to our final remarks, it is worth mentioning that it
would have been theoretically appealing to have FN-inspired textures
for both $y$ and $\tilde y$ couplings; as can be indirectly inferred
from the above discussion, we did not succeed in finding
phenomenologically viable $\tilde y$ couplings with textures
mirroring those of $y$.

\section{Concluding remarks}{\label{sec:conclusion}}
In this work we have carried a comprehensive phenomenological study of
a SM extension via two scalar leptoquarks $h_{1,2}$ and 
three generations of triplet neutrinos $\Sigma_R^i$, further reinforcing
the SM gauge group via a discrete $Z_2$ symmetry under which $h_2$ and
$\Sigma_R^i$ are odd (all other fields being even). 

The present New Physics construction aims at simultaneously addressing
two long-standing SM observational problems - neutrino mass
generation, and a viable dark matter candidate - while further
offering a solution to the currently reported anomalies in $B$ meson
decays, $R_{K^{(*)}}$. 

The $Z_2$ symmetry ensures the stability of the LZoP, rendering the
neutral component of the lightest $\Sigma_R$ a viable cold dark
matter candidate for well defined intervals of its mass. 
In the absence of a full theory of flavour, we have identified several
classes of flavour textures for the $h_1$ leptoquark Yukawa couplings
which succeed in saturating the $R_{K^{(*)}}$ anomalies. 
These textures (loosely based on Froggatt-Nielsen inspired ans\"atze)
were subjected to a vast array of flavour conserving and flavour
violating observables (including meson decays, neutral meson
oscillations and cLFV decays), which allowed to infer stringent
constraints on the $h_1$ leptoquark mass and couplings. 
Contrary to previous claims in the literature, our findings
suggest that the strongest constraints on these leptoquark
extensions do arise from cLFV $\mu -e$ conversion in nuclei, and 
from the rare $K^+ \to \pi^+ \nu \bar \nu$ decays. 
Furthermore, we also verified that numerous ans\"atze (identified as
promising ones for leptoquarks couplings, see
e.g.~\cite{Nomura:2016ezz,Cheung:2016frv}) were in fact phenomenologically 
disfavoured by several of the here considered observables. 

It is important to emphasise that the constraints on leptoquark
couplings arising from flavour observables are not intrinsic (nor
peculiar) to the leptoquark realisation here considered; in
fact these are valid for any SM extension via scalar triplet leptoquarks.

\bigskip
The present BSM realisation leads to a scenario in which neutrino
masses are radiatively generated (at the three-loop level, from the
exchange of leptoquarks, down-type quarks and lepton
triplets, see Fig.~\ref{fig:numass}). 
Neutrino oscillation data can be accounted for by means of
a modified Casas-Ibarra parametrisation: avoiding 
non-perturbative regimes for the Yukawa couplings, 
$y$ and $\tilde y$, establishes the final constraints on the parameter
space of the model.

\bigskip
The inclusion of Majorana states opens the door to lepton number
violating processes; the radiatively induced masses for the light
(left-handed) neutrinos are one such example. The new interactions and
couplings further allow for additional sources of CP violation. 
It is thus only natural to envisage the possibility of accounting for
the baryon asymmetry of the universe. 
In the present realisation, one can have
tree-level processes which are lepton number violating: one such
example can be obtained from the neutrino mass (loop) diagrams - see
Fig.~\ref{fig:numass} -, by ``cutting'' the inner fermion
propagators. This would lead to tree-level LNV decays of 
the heavier neutral  
$\Sigma^{2,0}_R$ into, for instance, $\Sigma^{1,0}_R + \bar d + \bar d + 
d\nu_L+d\nu_L$ (which could have CP violating interferences with
higher order diagrams). However, these appear to be heavily suppressed
processes and, in the absence of a detailed evaluation, it remains
unclear whether one could indeed generate a significant lepton
asymmetry. In addition,  
$\Sigma_R$ decoupling would be required to occur above the EW phase
transition to have an efficient conversion into a baryon asymmetry.

\bigskip
In summary, and following a thorough study of an extensive array of
observables, we have proposed realisations of a SM scalar leptoquark
extension capable of accommodating neutrino oscillation data, a
viable DM candidate, and saturating the observed discrepancies for 
$R_{K^{(*)}}$.
We notice that the present construction cannot account for the
tensions in $R_{D^{(*)}}$, nor for the discrepancy between observation
and SM prediction in what concerns the muon anomalous magnetic
moment. Should the latter persist, then the candidate model here
studied will have to be extended, or then embedded in a larger
framework \cite{Becirevic:2018afm}. 

In the near future, a number of high-intensity experiments will put
the present leptoquark construction to the test, in particular 
several cLFV-dedicated facilities (searches for radiative and
three-body muon decays, in addition to neutrinoless conversion in
nuclei) and a possible measurement of the rare decay $K^{+}\to
\pi^{+} \nu \bar{\nu}$. Hopefully, positive signals or new
stringent bounds emerging from negative searches, will allow to
further constrain the model's parameter space, or possibly disfavour
it as a candidate New Physics model.

\section*{Acknowledgements}
We are grateful to Damir~Be\v{c}irevi\'{c} and Gudrun Hiller
for valuable discussions and suggestions.
C.H., J.O. and A.M.T. acknowledge support within the framework of the
European Union's Horizon 2020 research and innovation programme under
the Marie Sklodowska-Curie grant agreements No 690575 and No 674896.

\end{document}